\documentclass[useAMS,usenatbib]{mn2e}
\usepackage[utf8]{inputenc}
\usepackage{graphicx}
\usepackage{amsmath}
\usepackage{amssymb}
\usepackage{times}
\usepackage{color}
\usepackage[usenames,dvipsnames,svgnames,table]{xcolor}
\usepackage{xspace}

\definecolor{customhdrcolor}{rgb}{0.0,0.0,0.0}
\definecolor{customcitecolor}{rgb}{0.0,0.5,0.75}
\definecolor{customlinkcolor}{rgb}{0.0,0.5,0.75}

\usepackage[colorlinks=true,linkcolor=customlinkcolor,urlcolor=customlinkcolor,citecolor=customcitecolor,pdftex]{hyperref}

\ifpdf\else\usepackage{graphics}\fi

\newcommand{\degree}{\ensuremath{^{\circ}}\xspace}

\DeclareRobustCommand{\TUSSEN}[3]{#2}

\title[EoR foregrounds and point-source spectra]{Parametrising Epoch of Reionization foregrounds: A deep survey of low-frequency point-source spectra with the MWA}

\def\ASTRON{$^{1}$}
\def\Curtin{$^{2}$}
\def\CAASTRO{$^{3}$}
\def\Wellington{$^{4}$}
\def\UMelbourne{$^{5}$}
\def\UW{$^{6}$}
\def\ASU{$^{7}$}
\def\ANU{$^{8}$}
\def\MIT{$^{9}$}
\def\Berkeley{$^{10}$}
\def\USydney{$^{11}$}
\def\Dunlap{$^{12}$}
\def\CfA{$^{13}$}
\def\CASS{$^{14}$}
\def\Brown{$^{15}$}
\def\RRI{$^{16}$}
\author[A.~R.~Offringa et al.]{A.~R.~Offringa\ASTRON\thanks{Corresponding author. E-mail: \url{offringa@astron.nl}},
C.~M.~Trott\Curtin$^,$\CAASTRO,
N.~Hurley-Walker\Curtin,
M.~Johnston-Hollitt\Wellington, \newauthor
B.~McKinley\UMelbourne$^,$\CAASTRO,
N.~Barry\UW,
A.~P.~Beardsley\ASU,
J.~D.~Bowman\ASU, 
F.~Briggs\ANU$^,$\CAASTRO, 
P.~Carroll\UW,\newauthor
J.~S.~Dillon\MIT$^,$\Berkeley, 
A.~Ewall-Wice\MIT,
L.~Feng\MIT, 
B.~M.~Gaensler\USydney$^,$\CAASTRO$^,$\Dunlap,
L.~J.~Greenhill\CfA, \newauthor
B.~J.~Hazelton\UW,
J.~N.~Hewitt\MIT, 
D.~C.~Jacobs\ASU, 
H.-S.~Kim\UMelbourne$^,$\CAASTRO, 
P.~Kittiwisit\ASU, 
E.~Lenc\USydney$^,$\CAASTRO, \newauthor
J.~Line\UMelbourne$^,$\CAASTRO,
A.~Loeb\CfA, 
D.~A.~Mitchell\CASS$^,$\CAASTRO, 
M.~F.~Morales\UW,
A.~R.~Neben\MIT,
S.~Paul\RRI,  \newauthor
B.~Pindor\UMelbourne$^,$\CAASTRO,
J.~C.~Pober\Brown, 
P.~Procopio\UMelbourne$^,$\CAASTRO, 
J.~Riding\UMelbourne$^,$\CAASTRO,
S.~K.~Sethi\RRI,
N.~U.~Shankar\RRI, \newauthor
R.~Subrahmanyan\RRI,
I.~S.~Sullivan\UW, 
M.~Tegmark\MIT, 
N.~Thyagarajan\ASU,
S.~J.~Tingay\Curtin$^,$\CAASTRO, \newauthor
R.~B.~Wayth\Curtin$^,$\CAASTRO,  
R.~L.~Webster\UMelbourne$^,$\CAASTRO, 
J.~S.~B.~Wyithe\UMelbourne$^,$\CAASTRO
\\
\ASTRON{}Netherlands Institute for Radio Astronomy (ASTRON), PO Box 2, 7990 AA Dwingeloo, The Netherlands\\
\Curtin{}International Centre for Radio Astronomy Research, Curtin University, Bentley, WA 6102, Australia\\
\CAASTRO{}ARC Centre of Excellence for All-sky Astrophysics (CAASTRO)\\
\Wellington{}School of Chemical and Physical Sciences, Victoria University of Wellington, PO Box 600, Wellington 6140, New Zealand\\
\UMelbourne{}School of Physics, University of Melbourne, Parkville, VIC 3010, Australia\\
\UW{}Department of Physics, University of Washington, Seattle, WA 98195, USA\\
\ASU{}School of Earth and Space Exploration, Arizona State University, Tempe, AZ 85287, USA\\
\ANU{}Research School of Astronomy and Astrophysics, Australian National University, Canberra, ACT 2611, Australia \\
\MIT{}Kavli Institute for Astrophysics and Space Research, Massachusetts Institute of Technology, Cambridge, MA 02139, USA\\
\Berkeley{}Berkeley Center for Cosmological Physics, University of California, Berkeley, Berkeley, CA 94720, USA\\
\USydney{}Sydney Institute for Astronomy, School of Physics, University of Sydney, NSW 2006, Australia\\
\Dunlap{}Dunlap Institute for Astronomy and Astrophysics, University of Toronto, Toronto, ON M5S 3H4, Canada\\
\CfA{}Harvard-Smithsonian Center for Astrophysics, Cambridge, MA 02138, USA\\
\CASS{}CSIRO Astronomy and Space Science, Marsfield, NSW 2122, Australia\\
\Brown{}Department of Physics, Brown University, Providence, RI 02912, USA.\\
\RRI{}Raman Research Institute, Bangalore 560080, India
}

\begin{document}

\date{Accepted 2016 February 5. Received 2016 February 4; in original form 2016 January 6.}
\pagerange{\pageref{firstpage}--\pageref{lastpage}}
\pubyear{2015}

\label{firstpage}
\maketitle

\begin{abstract}
Experiments that pursue detection of signals from the Epoch of Reionization (EoR) are relying on spectral smoothness of source spectra at low frequencies. This article empirically explores the effect of foreground spectra on EoR experiments by measuring high-resolution full-polarization spectra for the 586 brightest unresolved sources in one of the MWA EoR fields using 45~h of observation. A novel peeling scheme is used to subtract 2500 sources from the visibilities with ionospheric and beam corrections, resulting in the deepest, confusion-limited MWA image so far. The resulting spectra are found to be affected by instrumental effects, which limit the constraints that can be set on source-intrinsic spectral structure. The sensitivity and power-spectrum of the spectra are analysed, and it is found that the spectra of residuals are dominated by PSF sidelobes from nearby undeconvolved sources. We release a catalogue describing the spectral parameters for each measured source.
\end{abstract}


\section{Introduction}
The signature of the cosmological Epoch of Reionization (EoR) is directly detectable by the redshifted 21-cm HI line. Several EoR experiments are underway to detect this signature in low-frequency observations, which will potentially result in a better understanding of this important epoch. These experiments either aim to detect spectral fluctuations in the global signal using a single element (\citealt{edges-2010,burns-2012-dare,voytek-2014-scihi,bighorns-2015}; \citealt*{bernardi-2015-leda}), or to detect spectral and spatial variations using an interferometer, such as with GMRT \citep{paciga-2013-GMRT-EoR}, LOFAR \citep{ncp-eor-yatawatta}, MWA \citep{bowman-science-with-the-mwa-2013} and PAPER \citep{ali-eor-paper-2015}.

This work uses the Murchison Widefield Array (MWA; \citealt{mwa-2013,bowman-science-with-the-mwa-2013}) to analyse the spectral characteristics of discrete foreground sources and instrumental effects that affect these foreground spectra. An early result with the 32-tile MWA prototype reached an upper limit for the EoR signals of $\Delta^2(k)= 9 \times 10^4$ mK$^2$ at a comoving scale $k = 0.046$ Mpc$^{-1}$ and $z = 9.5$ after 22 h of observing \citep{mwa32-eor-limit-2014}. It has been theoretically shown that the 128-tile MWA can perform a significant detection of the EoR signal in one field after integrating 1000 hours, assuming ideal foreground subtraction \citep{mwa-eor-sensitivity-2013,thyagarajan-2013-mwa-eor-foregrounds}. A first analysis with the full 128-tile MWA, using 3 h of integration time has reached a limit of $\Delta^2(k)= 3.7 \times 10^4$ mK$^2$ at $k=0.18$ h Mpc$^{-1}$ \citep{mwa-eor-limit-2015}. Assuming further integration does not reveal any systematic effects, this implies an integration time of 3000~h is required for a detection of the expected signals of $\sim$10mK$^2$. The cause of the difference between the theoretical and practical required integration time is being investigated. The most competitive upper limit for the EoR signal is currently $\Delta^2(k)=5.0\times 10^2$ mK$^2$ at $z$=8.4 and $k=0.15$ $h$Mpc$^{-1}$, which has been achieved using PAPER \citep{ali-eor-paper-2015}.

Detecting the EoR signal is a challenging task. Apart from the requirement of long integration times, foreground sources are orders of magnitude brighter than the EoR signal. While the EoR signal is expected to have small-scale (unsmooth) spectral features, astrophysical sources are dominated by synchrotron emission at low frequencies, and have sufficiently smooth spectra to separate them from the EoR signal. \citet*{datta-2010-eor-foreground-subtraction} first identified that a two-dimensional ($k_\parallel$,$k_\perp$) power spectrum would isolate power from smooth foregrounds in a ``wedge'' area. Others have subsequently explored the origin of this wedge (\citealt*{vedantham-eor-foregrounds-2012}; \citealt*{trott-2012-eor-point-sources}; \citealt{morales-2012-eorwindow}; \citealt{parsons-2012-delay-spectrum}; \citealt{thyagarajan-2013-mwa-eor-foregrounds}). This foreground behaviour makes it possible to distinguish them from the EoR signal. Sharp spectral features that are known to exist at low frequencies, such as radio-recombination lines \citep{asgekar-lofar-rrls-cas-a-2013,morabito-2014} and high-redshift HI absorption \citep{ciardi-2012-21cm-forest}, are sufficiently weak not to be an issue. Therefore, it is generally assumed that foreground sources can be modelled with smooth functions, such as double-logarithmic polynomials of low order \citep{wang-2006,mcquinn-2006,jelic-lofar-foregrounds-2008,liu-2009}. Polarized sources are a possible concern, because they can introduce artefacts into total intensity spectra (e.g. \citealt*{geil-2011-pol-foregrounds}).

A few studies have focused on the spatial behaviour of low-frequency 21-cm foregrounds (\citealt*{ali-gmrt-foregrounds-2008}; \citealt{bernardi-wsrt-foregrounds-II-2010,thyagarajan-2015-widefield-effect}), and its polarization \citep{jelic-lofar-foregrounds-II-2014,asad-lofar-polarization-leakage-2015}. However, the exact spectral behaviour of these foregrounds is mostly an unexplored area. An analysis of the frequency behaviour of 21-cm-foreground point-source spectra was performed with the GMRT at 150~MHz \citep{ghosh-21cm-foregrounds-2012}, and showed oscillations and unexplained curvature over frequency in the measured power spectra. Surveys such as MWACS \citep{hurley-walker-mwacs-2014} and MSSS \citep{heald-2015-msss} provide measurements of the spectrum of many sources at the redshifted EoR frequency, but their data points are integrated over large bandwidths, and do not provide information of the behaviour of sources and the instruments at high-resolution ($\Delta \nu < 250$ kHz). The MWA has recently been used to search for SH molecular lines at a resolution of 10 KHz in the Galactic centre \citep{tremblay-2015-mwa-molecular-lines}, which demonstrates the ability of the MWA to do spectral work at low frequency.

In this paper, we will perform a detailed study of spectra with high sensitivity and high frequency resolution for extra-galactic point sources. Thereby, we aim to assess both the ability to obtain sensitive spectra with the MWA, and to find if there are sources that have unexpected spectra that would be problematic for the EoR signal extraction.

\section{Observations \& Methods}
In the following sections, we describe the relevant MWA observations and the methods which we have used to reduce these data.

\subsection{Observations}
\begin{table}%
\caption{Observation nights used in the analyses. As indicated, three nights are not used because they do not calibrate well. The `band' column specifies whether the low 138.9--169.6~MHz band or high 167.0--197.7~MHz band is observed. The `snapshot' column specifies the number of 112-s snapshots that are usable. The `RMS' column specifies the residual RMS per 40~kHz spectral channel, after subtracting the best fitting model of the source with the lowest RMS.}%
\label{tbl:data}\begin{center}\begin{tabular}{l|c|l|r|r|c}%
\textbf{Date} & \textbf{Used?} & \textbf{Band} & \textbf{Res.} & \textbf{Snapshots} & \textbf{RMS} \\
  \hline
  2013-08-23 & yes & High & $2.0'$ & 132 & 50.4 mJy \\
  2013-08-26 & yes & Low  & $2.3'$ & 143 & 74.7 mJy \\
  2013-09-12 & yes & Low  & $2.3'$ & 143 & 64.9 mJy \\
  2013-09-13 & yes & High & $2.0'$ & 132 & 51.0 mJy \\
  2013-09-17 & yes & High & $2.0'$ & 132 & 46.3 mJy \\
  2013-09-18 & yes & Low  & $2.3'$ & 143 & 60.7 mJy \\
  2013-09-19 & yes & High & $2.0'$ & 132 & 51.3 mJy \\
  2013-09-20 &  no & Low  & $2.3'$ & 142 & --- \\
  2013-09-30 &  no & High & $2.0'$ & 142 & 63.2 mJy \\
  2013-10-01 &  no & Low  & $2.3'$ & 143 & 64.9 mJy \\
  2013-10-02 & yes & High & $2.0'$ &  70 & 61.0 mJy \\
  2013-10-03 & yes & Low  & $2.3'$ &  89 & 75.6 mJy \\
  2013-10-09 & yes & Low  & $2.3'$ & 113 & 66.7 mJy \\
  2013-10-10 & yes & High & $2.0'$ & 113 & 52.9 mJy \\
  2013-10-11 & yes & Low  & $2.3'$ & 113 & 79.6 mJy \\
  \hline
   \multicolumn{4}{l|}{12/15 nights used} & 1481/1898 & 32.6 mJy
 \end{tabular}  
 \end{center}
\end{table}

The observations used in this work have been made as part of the MWA EoR project. We have used observations that are centred at RA 0\degree, Dec -27\degree, and recorded between August and October of 2013. The field around this target is referred to as the MWA EoR0 field -- one of three fields that were selected based on having weak Galactic foregrounds and passing nearly through zenith at the MWA.

The selected 15 nights are listed in Table~\ref{tbl:data}. Of these 15 nights, 3 nights were not included in the analyses because they show RFI or unusual calibration solutions. The MWA can observe 30.72~MHz simultaneously. To cover a larger redshift range, a total bandwidth of 138.9--197.7~MHz is recorded by observing in two different bands. The low band covers 138.9--169.6~MHz and the high band covers 167.0--197.7~MHz. Together these cover the HI 21-cm line at redshifts 6.1--9.2. The observations have a frequency resolution of 40~kHz and time resolution of 0.5~s.

A pointing procedure is used in which the electronically-steered pointing direction of the telescope is kept constant for a while, typically about 30 min, thereby letting the field drift through the primary beam, before the telescope is repointed to track the target field. This is because the antenna delays are restricted to a certain quantization. The pointing directions that are chosen with this procedure provide an optimized sensitivity.

\subsection{Data analysis}
In this section we will describe the data processing strategy required to extract the images and source spectra from the data. Our data processing strategy includes several novel methods and tools, and we will therefore describe these in detail.

The first steps in our data processing are to flag RFI, average the data in time to 4~s and convert the raw data to measurement sets. A time resolution of 4~s is high enough to prevent decorrelation up to the first null of the primary beam. These steps are performed by the \textsc{cotter} preprocessing pipeline \citep{offringa-2015-mwa-radio-environment}, which uses an \textsc{aoflagger} strategy for RFI detection (\citealt{offringa-2010-post-correlation-rfi-classification}; \citealt*{offringa-2012-scale-invariant-rank-operator}) that was optimized for the MWA.

Each night is split up in snapshots of 112~s, and each snapshot is globally calibrated using a source model in which the spectral energy distribution of each source is assumed to follow a power law. The spectral index in the model is independent for each source. The model is bootstrapped from cross-matching the MWA commissioning survey \citep{hurley-walker-mwacs-2014} at 180 MHz to the SUMMS catalogue at 843 MHz \citep{summs-II-2003}. The observation contains a few clearly resolved sources, most prominently nearby galaxies in the Sculptor group (see Sect.~\ref{resolved}). Such sources are found by hand and subsequently modelled with multiple point components. Fainter sources are added to the calibration model after a first imaging iteration of two nights. These sources are given a power law formed from their measured flux density combined with a measurement from other catalogues covering the source. For this, also the 408-MHz Molonglo Reference Catalogue (MRC, \citealt{mrc-1981}) is used. If no second flux density measurement is available, the source is assigned to follow a power law formed from the low and high band observations. Source detection is performed with the \textsc{aegean} source finder \citep{aegean-hancock-2012}. The end result is a model with $\sim$16,000 sources in an area of $45\degree\times45\degree$, all with independent spectral indices.

The first calibration is performed as a direction-independent full-polarization self-calibration. This is performed with the \textsc{mitchcal} tool, which is the authors' custom implementation of the algorithm described by \citet{rts-mwa-2008}\footnote{This algorithm was later rediscovered by Stefano Salvini and subsequently named \textsc{stefcal} \citep{stefcal-2014}.}. Each 40-kHz channel is independently calibrated. After global calibration, a few thousand sources are peeled using a clustered peeling procedure that mitigates the ionosphere by fitting positions and gains in 25 directions, which are the centres of the 25 clusters. Clusters were made by using an angular k-means clustering algorithm to group the modelled sources, as described by \citet*{kazemi-clustered-calibration-2013}. The peeling was performed by a tool named \textsc{ionpeel}, which was also specifically written for the MWA. For each cluster of sources, it performs a Levenberg-Marquardt (LM) least-squares optimisation between model and data for the parameters $\Delta l, \Delta m$ and $g$, being the $l$ and $m$ position offsets and the gain factor. After a solution is found for a cluster, the cluster is subtracted from the data with the current best $\Delta l, \Delta m$ and $g$, and this procedure is repeated 3 times for all clusters to minimize the effect that clusters have on each other. An independent fit for these three parameters is performed per cluster for every 4~channels (160~kHz).

For quality assurance, the peeled snapshots are imaged using \textsc{wsclean} \citep{offringa-wsclean-2014} on a $5120\times5120$ image of $30''\times30''$ pixels with uniform weighting. The resolution of the MWA is $2.3'$ at these frequencies. Snapshots with deviating image noise levels are removed from further analyses. Because sources have already been peeled, some deconvolution has already been performed, but further deconvolution is performed by cleaning each snapshot to 100~mJy. The noise RMS in an average snapshot is 25~mJy/beam. To create the final integrated images, the peeled sources are restored and the images are corrected for the MWA beam model and weighted accordingly, before they are added together. The Jones matrices of the beam are calculated by electromagnetic simulations of the tiles as described by \citet{mwa-beam-sutinjo-2015}. Beam corrections are applied to the linearly-polarized images, by inverting the beam voltage matrix $B$ for each pixel's polarization matrix $I$, and computing $B^{-1}IB^{*-1}$, where $^{*}$ denotes the conjugate transpose, as described in \citet{offringa-wsclean-2014}.

\begin{figure*}
\begin{center}
\includegraphics[width=18cm]{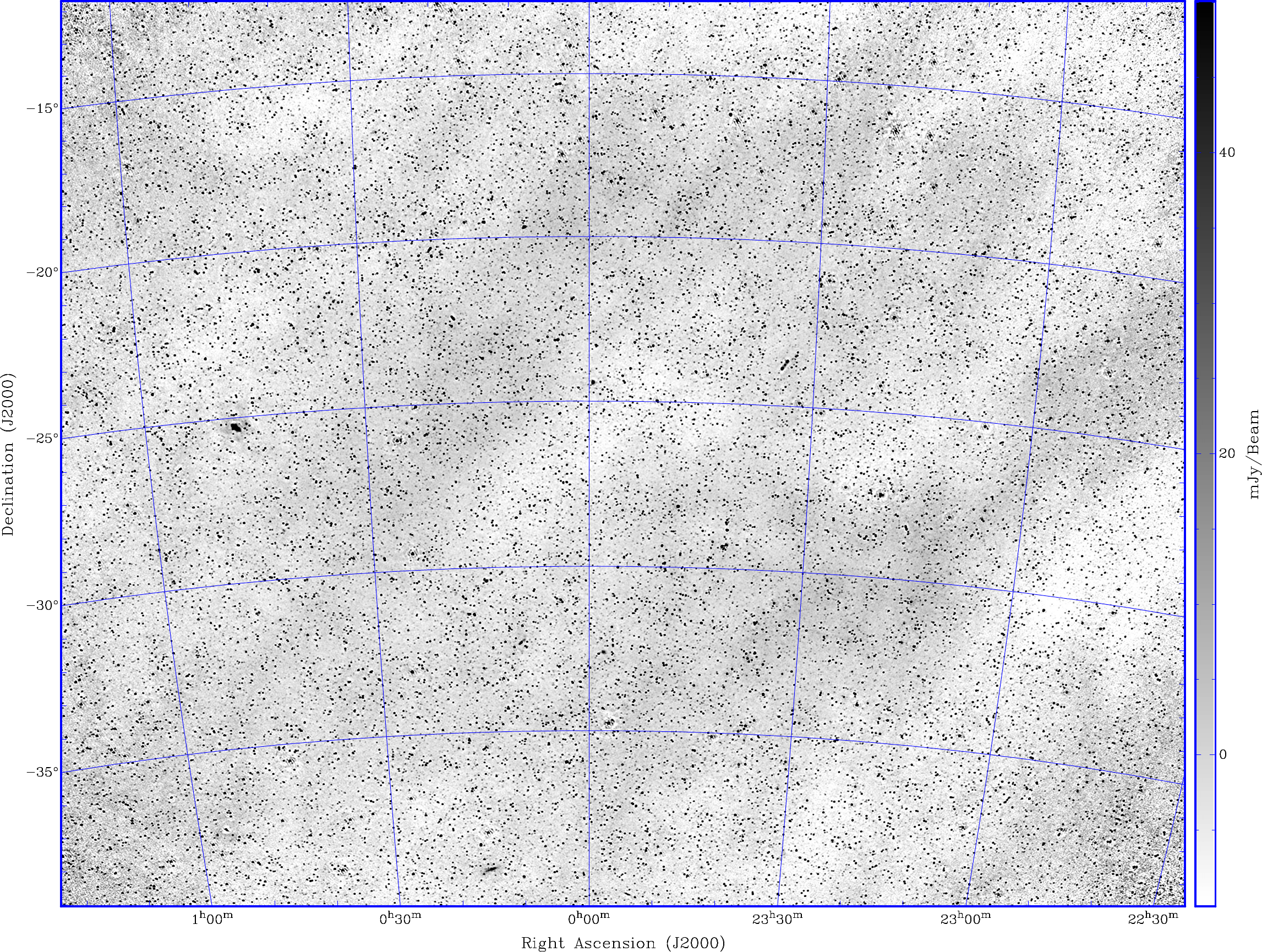}
\caption{$\sim$45$\degree \times 30\degree$ beam-corrected map of the EoR0 field, after 45 hours of integration and averaging the low and high bands together. Noise becomes apparent in the corner of the image due to the primary beam null. }
\label{fig:eor0-fullbw-30degcut}
\end{center}
\end{figure*}

Finally, the point-source spectra are determined from the peeled visibilities, by measuring the flux density at the positions that were found during peeling, weighted with the beam. This is performed by calculating the direct inverse Fourier transform. Our final estimate for the spectra $S(\nu)$ are given by the sum of the peeled flux densities and residuals, $\hat{S} (\nu) = S^\nu_\textrm{peel} + S^\nu_\textrm{res}$. Both of these are beam-corrected $2\times2$ matrices containing the linearly-polarized flux densities, i.e., the \texttt{xx}, \texttt{xy}, \texttt{yx} and \texttt{yy} correlations. $S_\textrm{res}$ and $S_\textrm{peel}$ are calculated with
\begin{equation}
 S^\nu_\textrm{res} = %
   \left( \sum \limits_{j \in \Upsilon_\nu} \gamma_j B^*_j V_j e^{2 \pi i \left[u_j \tilde{l}_j + v_j \tilde{m}_j + w_j(\sqrt{1 - \tilde{l}_j^2 - \tilde{m}_j^2}-1) \right] } B_j \right) W^{-1}
\end{equation}
and
\begin{equation}
 S^\nu_\textrm{peel} = %
   \left( \sum \limits_{j \in \Upsilon_\nu} \gamma_j g_j B^*_j B_j M B_j^* B_j \right) W^{-1},
\end{equation}
where $W$ is the $2\times 2$ normalization matrix,
\begin{equation}
 W = \sum \limits_{j \in \Upsilon_\nu} \gamma_j B^*_j B_j B^*_j B_j.
\end{equation}
Here, $\Upsilon_\nu$ is a set with indices that select the visibility matrices at frequency $\nu$ over which the summation is performed; $V$ is a $2\times2$ visibility matrix; $B$ is the beam Jones matrix at the (uncorrected) position of the source at the time and frequency of the corresponding visibility; $\gamma$ is the weight of the visibility matrix (determined from the `WEIGHT\_SPECTRUM' column of the measurement set), $u, v$ and $w$ represent the visibility baseline coordinates, $(\tilde{l},\tilde{m})$ is the corrected source position ($\tilde{l}=l+\Delta l$), and $M$ is the absolute model flux density matrix of the source (such that $g M$ is the flux density found during peeling).

These equations are such that an incorrect model or invalid peeled gain value $g$ do not influence the found flux density value, because an invalid model and/or invalid gain $g$ will leave more residuals behind, and this cancels out when adding together $S_\textrm{res}$ with $S_\textrm{peel}$. This is of course important, because we do not want to enforce the power laws from our model onto the measured spectra. Peeling influences only the position at which the flux density is determined, and performs the deconvolution. When peeling a cluster results in divergence, the involved visibilities are excluded from the computation.

This method evaluates and applies the beam correctly for each timestep, channel and source position. Because the MWA beam was only modelled at 1.2 MHz frequency intervals \citep{mwa-beam-sutinjo-2015}, the beam values are interpolated to 40~kHz using spline interpolations.

The above equations are evaluated for all peeled point sources. Extended sources (those with multiple components in the model) are not measured. To be able to get `cleaned' spectra, it is assumed that the peeling procedure has deconvolved the data. While the residual images after peeling are indeed reasonably empty, some sources are still visible, because of subtraction errors and an incomplete model. The faint diffuse Galactic synchrotron radiation has also not been deconvolved. The flux density resulting from the above equations is very sensitive to PSF sidelobes, because visibilities are weighted with a natural scheme, and initial results from two nights showed large-scale oscillations going through the spectra. This was found to be caused by insufficient deconvolution, either from residual point sources or from Galactic diffuse emission. Therefore, the values were recalculated with a uniform weighting scheme; each visibility weight $\gamma_j$ was additionally multiplied with a weight determined from binning the $u,v,w$ positions, in the same way as is done for uniform imaging. This procedure increases the noise in the spectra, but greatly decreases the effect of imperfect deconvolution.

Peeling and spectrum extraction are the most expensive tasks during the processing, despite that these tasks are implemented in a multi-threaded way. Because of the computational cost of these operations, we have chosen to peel and measure only the 2500 brightest sources of our total 16,000 source catalogue. With 2500 sources, both of the operations take several hours on a single 112~s snapshot. Using the Australian National Computational Infrastructure (NCI) cluster `Raijin', we were able to run these operations on approximately 100 nodes at a time. Using these 100 nodes, processing a full night of observations takes approximately one day, which consists of \textsc{cotter} preprocessing, global calibration with \textsc{mitchcal}, imaging with \textsc{wsclean}, peeling with \textsc{ionpeel} and extracting the spectra.

\section{Results}

\subsection{Imaging results} \label{sec:imaging-results}
\begin{figure}
\begin{center}
\includegraphics[width=8cm]{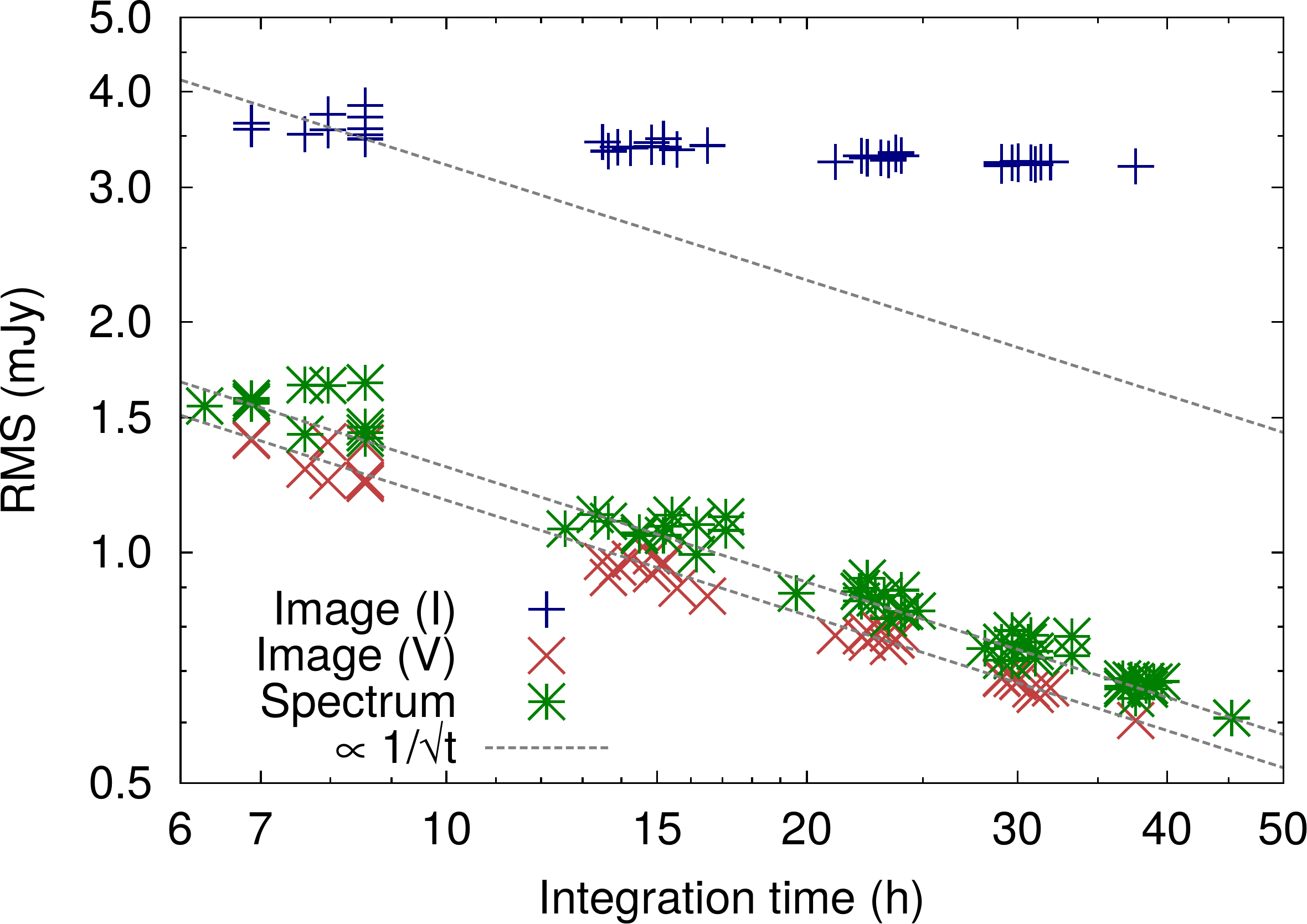}
\caption{Sensitivity as function of time for the images integrated over the total 59~MHz bandwidth and for the spectrum measured at 40~kHz. The spectral standard deviations (calculated as described in \S\ref{sec:spectra-sensitivity}) are converted to the equivalent total-bandwidth noise. After 5~$h$ of integration, the standard deviation of the total-power imaging noise continues to decrease slightly when increasing the integration time, but not proportionally to $1/\sqrt{t}$ : the system noise does no longer contribute to the imaging noise at this point, except on the longest baselines. Therefore, the classical and/or sidelobe confusion level is approximately reached. The Stokes~V imaging noise and the spectrum RMS do behave like system noise.}
\label{fig:combined-sensitivity-over-time}
\end{center}
\end{figure}
\begin{table*}%
\caption{Comparison of the flux density measurements of the three brightest sources in the field of view.}%
\label{tbl:fluxcomparison}\begin{center}\begin{tabular}{l|c|c|l|c|c|c}%
 \multicolumn{3}{c}{}                            & \multicolumn{4}{c}{\textbf{Flux density measurements}} \\
\textbf{NED Name} & \textbf{RA} & \textbf{Dec} & \textbf{This work 168 MHz} & \textbf{MWACS 180 MHz} & \textbf{Culgoora 160 MHz} & \textbf{PAPER 145 MHz} \\
\hline
 ESO 349-G010  & 23h57m00.7s & -34d45m31.7s & 21.8 Jy & 21.8 Jy & 25.3 Jy & 13.9 Jy \\
 PKS 0023-26     & 00h25m49.2s & -26d02m12.8s & 21.2 Jy & 21.4 Jy & 20.8 Jy & 8.6 Jy \\
 PKS 0021-29     & 00h24m30.1s & -29d28m48.9s & 17.4 Jy & 17.0 Jy & 18.4 Jy & 15.1 Jy \\
\end{tabular}\end{center}\end{table*}
While the focus of this study is on the spectral behaviour of the foregrounds and instrument, we briefly analyse the images to study the imaging noise behaviour and possible instrumental artefacts in image space. The deep catalogue that results from the processing of these data will be described in later papers.

Fig.~\ref{fig:eor0-fullbw-30degcut} shows the map of all data of both bands after peeling. It is the beam-corrected and beam-weighted average of all the restored snapshots. Peeled sources are restored as Gaussians. Some deconvolution artefacts are visible at a level of a few mJy/beam, which are due to unmodelled and therefore unpeeled sources and insufficient cleaning (this is most easily seen in the corners of Fig.~\ref{fig:eor0-fullbw-30degcut}). The latter is because snapshots are cleaned separately, with a cleaning threshold of 100~mJy to avoid selecting noise peaks. Besides insufficient deconvolution, sources close to the beam null show some additional rippling artefacts of $\sim$10 mJy/beam (most visible in far top-left corner of Fig.~\ref{fig:eor0-fullbw-30degcut}). After further analysis it turned out these are caused by tile position errors. Such errors are absorbed in the calibration for sources in the centre. Flagging the worst offending tiles indeed attenuates these artefacts, but this was not yet done during the processing.

\textsc{aegean}'s ``\texttt{BANE}'' tool \citep{aegean-hancock-2012} separates foregrounds from noise and background, and calculates a mean noise level (image RMS) of $3.2\pm0.6$ mJy/beam over the central 10\degree of the image, which make it the deepest MWA image so far. Separate analysis of the two bands yields $3.6\pm0.7$ mJy/beam for the high band (with $22.1$ h of integration) and $4.4\pm0.8$ mJy/beam for the low band ($23.1$ h). While MWA's antenna response is optimized for the lower band (150~MHz), higher noise levels are observed in the lower band due to the increased sky noise at lower frequencies. The diffuse structure which can be seen in the image is Galactic emission. Since the low and high band images have the same diffuse structure, it is real emission and not sidelobe structure. The \textsc{aegean} source detector detects $30,027$ sources at $\ge5\sigma$ confidence in the full image.

To assess whether the image is confusion limited from either classical confusion or sidelobe confusion, we sample random combinations of nights (without replacement) and measure the noise of the integrated image using \texttt{BANE}. The results are in Fig.~\ref{fig:combined-sensitivity-over-time}. The total power images are approximately confusion limited after a single night, possibly less. The sensitivity keeps slightly increasing because the MWA only contains a few long baselines, causing a contribution of system noise to the smallest scales. The Stokes~V images are void of sources, except for weak sources that appear because of instrumental leakage. The Stokes~V leakage is typically $0.1$--$1$\% of the total brightness, and a visual inspection of the integrated Stokes~V image does not reveal any polarized sources that are distinguishable from the leakage. Because of the low brightness of sources in the Stokes~V image, its noise level continues to follow $1/\sqrt{t}$, as shown in Fig.~\ref{fig:combined-sensitivity-over-time}. The final Stokes~V image has an RMS of 0.6~mJy/beam.

A source population study using a single night of the low frequency band with equal processing strategy was performed by \citet{franzen-2015-mwa-source-population}. They find that the image is affected by sidelobe confusion noise at a $\sim 3.5$~mJy/beam level, and estimate the classical confusion limit at 154~MHz to be 1.7~mJy/beam. They also show that the measured source population down to 40~mJy in the MWA images is consistent with previous studies using the GMRT at the same frequency (\citealt{garn-2007-gmrt-eg-survey,intema-gmrt-bootes-I-2011,ghosh-21cm-foregrounds-2012}; \citealt*{williams-gmrt-mini-survey-2013}). \citet{franzen-2015-mwa-source-population} conclude that the flux scale in the MWA image is consistent with the GMRT studies.

As an example of the accuracy of the flux scale, we compare the three brightest sources in the field to other catalogues at the same frequency. Table~\ref{tbl:fluxcomparison} shows the following flux density measurements: this work; the MWA commissioning survey (MWACS; \citealt{hurley-walker-mwacs-2014}); the Culgoora catalogue \citep{calgoora-1995}; and the PAPER catalogue \citep{jacobs-2011-paper-survey}. Our measurements are consistent with MWACS and Culgoora assuming a 10\% error margin on both catalogues and our measurements. One of the PAPER catalogue measurements deviates more than 100\% from ours, but \citet{jacobs-2011-paper-survey} quote a 50\% standard error for their catalogue, and is likely therefore the cause of the deviation.

\begin{figure*}
\begin{center}
\includegraphics[width=8cm]{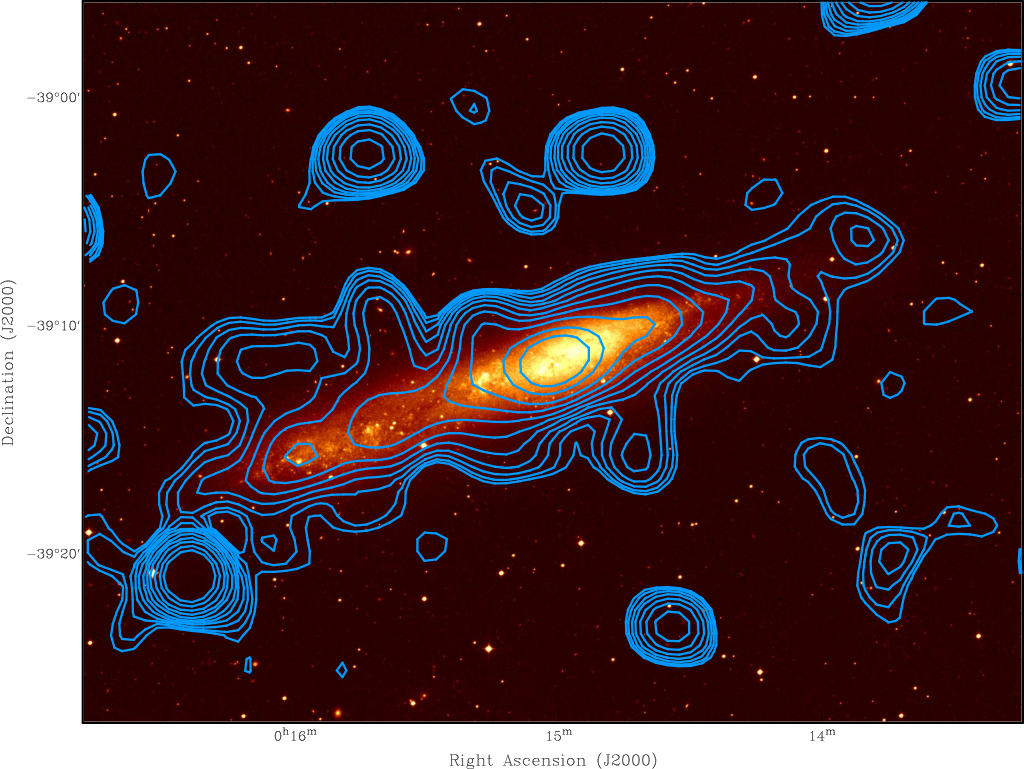}\hspace{5mm}\includegraphics[width=8cm]{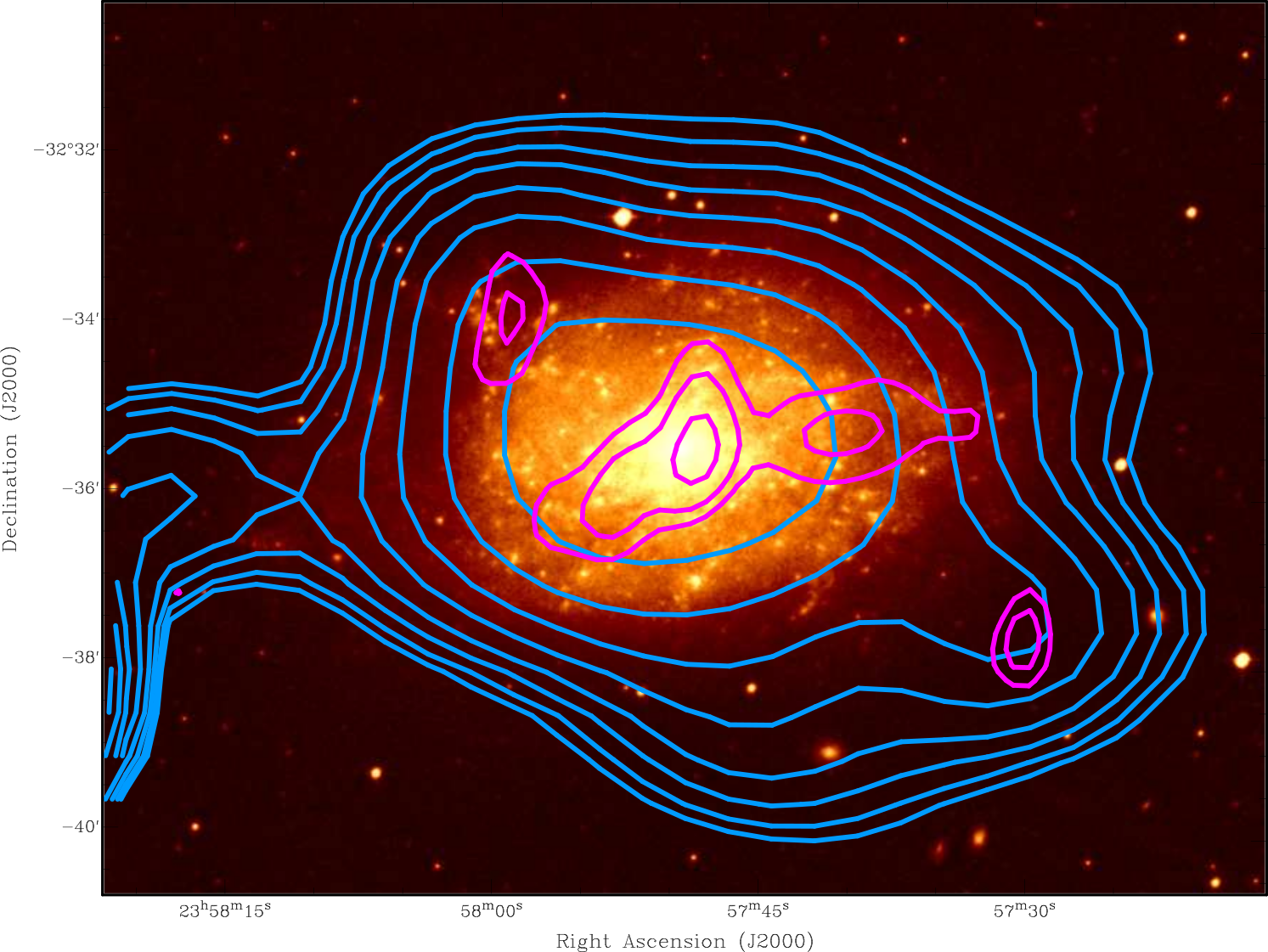}
\caption{HST images of NGC~55 (left) and NGC~7793 (right). Blue contour lines show
the MWA 168 MHz image from this work with contours starting at 5~mJy/beam
and increasing in intervals of $\sqrt{2}$. Magenta contour lines show the
SUMSS (843 MHz) in the right image with contours starting at 6~mJy/beam and
increasing in intervals of $\sqrt{2}$, the discrete source to the Northeast (RA 23:58:00, Dec. -32:34:00) appears to be associated with the micro-quasar S26.}
\label{fig:NGCs}
\end{center}
\end{figure*}%

\subsection{Resolved sources} \label{resolved}
In addition to the unresolved sources used in the spectral analysis there are a number of extended sources in the EoR0 field. A full characterisation of all of the extended sources will be presented elsewhere, and here we merely concentrated on the largest and most extended emission which required modelling with multiple components. The EoR0 field covers the region of the Sculptor Group of galaxies, which is a loose conglomeration of approximately 12 galaxies that has its centre only 3.9 Mpc from the Milky Way \citep{Karachentsev05} and is the closest group to the Local Group. Foremost among the group members is the so-called `Sculptor Galaxy', NGC~253, which is one of the brightest spirals beyond the Local Group, with a visual magnitude of 7.1. Radio emission across the disk and core of NGC~253 has been imaged by a variety of instruments over the preceeding 30 years covering a frequency range from 330~MHz to several GHz \citep{Turner85,Carilli96,Ulvestad97,Tingay04, Heesen05, Lenc06, Heesen11, rampadarath-2014-ngc253}. These data have revealed both the smooth, extended emission in the galactic disk and discrete sources contained within it. NGC~253 can be seen in Fig.~\ref{fig:eor0-fullbw-30degcut} (RA 00:47, Dec. -25:17) as the most prominent extended source in the field. Further details of the MWA emission of this source will be presented in a future publication (Kapinska et al. in prep). 

In addition to NGC~253, a number of other less studied Sculptor galaxies are strongly detected in the MWA image. In particular, NGC~7793 and NGC~55, shown in Fig.~\ref{fig:NGCs}, are both prominent extended sources in this deep 168~MHz image. NGC~55 is a Magellanic type, barred spiral galaxy which has been extensively observed in the optical, but little in the radio. In addition to the disk emission which spans 40', the MWA observations show a spur extending out of the plane of the galaxy, similar to the North polar spur feature in the Milky Way. NGC~7793 is classified as a chaotic spiral galaxy and is another prominent member of the Sculptor group notable due to the presence of a number of compact sources, including supernova remnants \citep{Pannuti02} and the microquasar S26 which hosts a blackhole of less than 15 solar masses \citep{Motch14}. No previous datasets have explored the disk emission, though there is some evidence of this in the archival SUMSS images (see Fig.~\ref{fig:NGCs}). The MWA observations clearly detect the low surface brightness emission across the entire disk of the galaxy. The MWA results for NGC~7793 and NGC~55 will be discussed elsewhere (Kapinska et al. in prep).

\subsection{Spectra}\label{sec:spectra}
\begin{figure}
\begin{center}
\includegraphics[width=8.3cm]{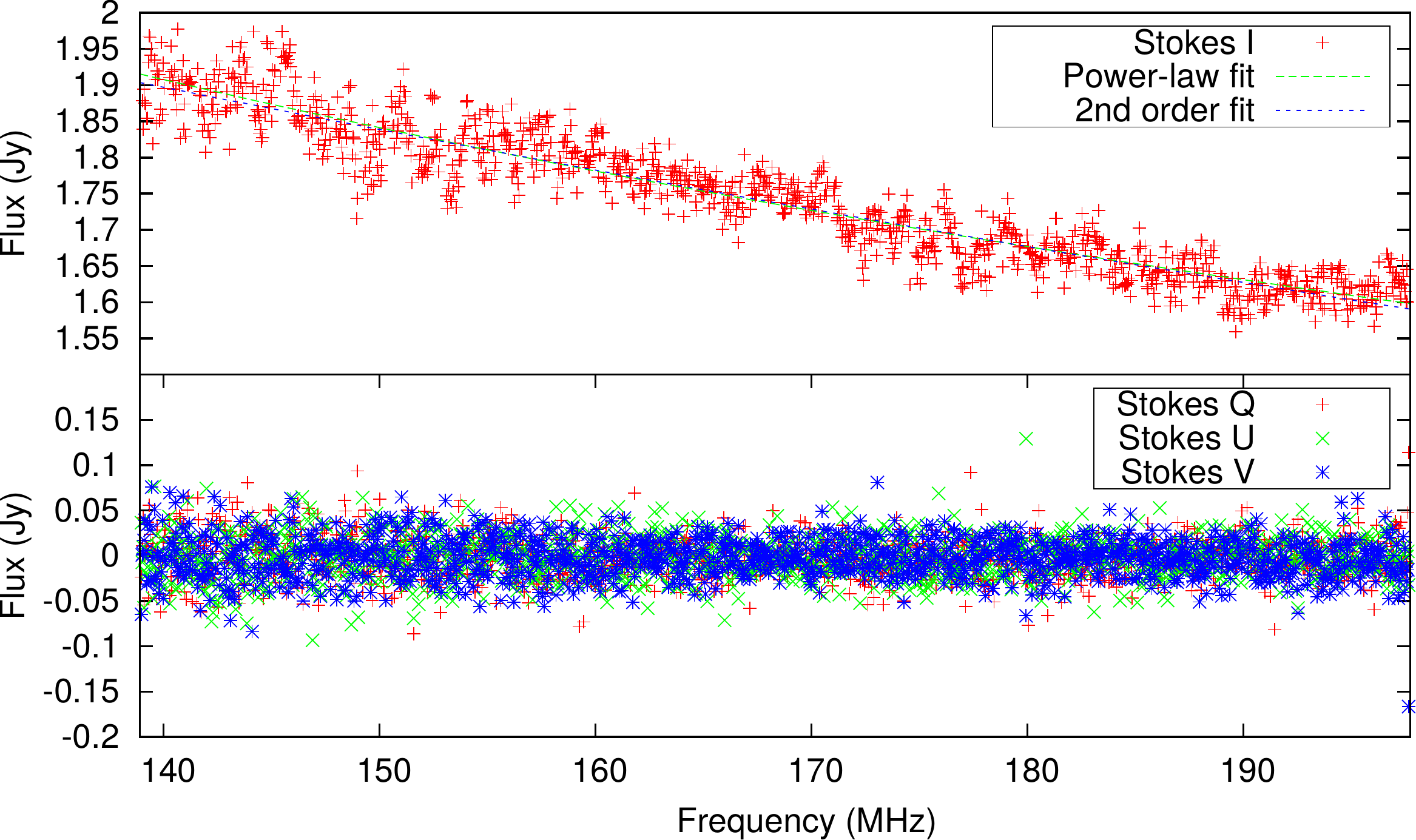}
\caption{One of the typical spectra produced in this work. The spectrum of this source (MWAEOR\,J000218-253915) has a residual RMS of 32 mJy. Some instrumental artefacts (see \S\ref{sec:cause-of-artefacts}) are visible as correlated structures in Stokes I. The other Stokes parameters are noise like. }
\label{fig:example-spectrum}
\end{center}
\end{figure}
\subsubsection{Sensitivity of spectra}\label{sec:spectra-sensitivity}
To assess whether the sensitivity in the spectra continues to increase when increasing integration time, we calculate the RMS for the source with the lowest RMS. To minimize the contribution from the signal, the RMS is calculated by differencing all adjacent channels and is divided by $\sqrt{2}$ to estimate the sensitivity in a single channel. The RMS is measured over the full bandwidth (138.9--197.7~MHz) at 40~kHz resolution, and converted to an equivalent bandwidth-integrated imaging noise level by dividing the RMS by $\sqrt{1242}$, where 1242 is the number of remaining (unflagged) channels. Some channels are flagged because of the poly-phase filter of the MWA, which divides the 30~MHz bandwidth in sub-bands of 1.28~MHz. The edges of each sub-band are contaminated by aliasing, and the central channel of each sub-band is lost due to the method of digitization of the signals in the MWA.

The integration time was varied by averaging a number of randomly selected nights as described in \S\ref{sec:imaging-results}, and each night is inverse-variance weighted before averaging. Each measurement contains the same number of low and high-band nights, so that the integration time is approximately constant over the full bandwidth. Fig.~\ref{fig:combined-sensitivity-over-time} shows the result. The spectral RMS $\propto 1/\sqrt{t}$, but the SED equivalent-noise level is on average 7\% higher than the Stokes~V imaging noise level. This increase could be due to the different processing strategy or due to systematics.

An example spectrum is shown in Fig.~\ref{fig:example-spectrum}, which shows the Stokes~I, Q, U and V spectra for source MWAEOR J000218-253915. The source has a fitted flux density of $1.74\pm0.10$~Jy. The polarized spectra are noise like, but the Stokes~I spectrum shows some artefacts. Similar artefacts are apparent in many spectra. We will analyse the cause and effect of these in later sections.

\subsubsection{Spectral indices}
\begin{figure}
\begin{center}
\includegraphics[width=8.3cm]{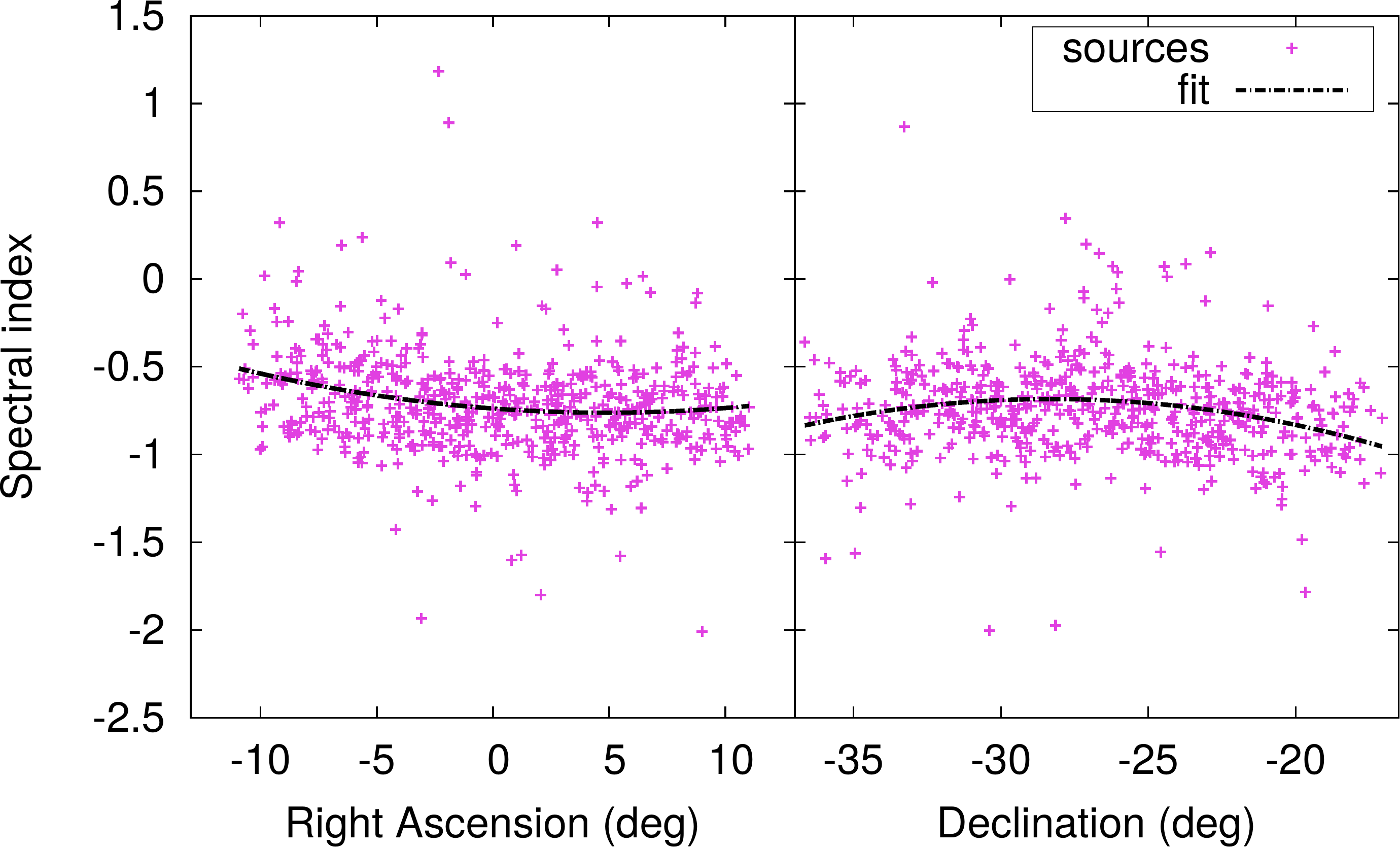}
\caption{Measured spectral indices fitted with a quadratic function, using a 10$\degree$ radius around the field centre (RA 0 h, dec -27$\degree$). The position-dependent flux-density bias shown by these plots are caused by errors in the beam model. }
\label{fig:si-vs-pos-fits}
\end{center}
\end{figure}

\begin{table}%
\caption{SI statistics, logarithmically binned by flux density. Columns show: flux density range of the bin; source count; average SI; median SI; and SI standard deviation.}%
\label{tbl:si-bined}\begin{center}\begin{tabular}{c|r|r|r|r}%
\multicolumn{1}{l|}{\textbf{Bin range}} & \textbf{N} & \textbf{$\mu_\textrm{SI}$} & \textbf{Med$_\textrm{SI}$} & \textbf{$\sigma_\textrm{SI}$} \\
  \hline
$\left[ 0.201 ; 0.648 \right>$ Jy & 286 & -0.671 & -0.683 & 0.288 \\ 
$\left[ 0.648 ; 2.09\right>$ Jy   & 225 & -0.726 & -0.747 & 0.276 \\ 
 $\left[ 2.09 ; 6.76\right>$ Jy   &  64 & -0.777 & -0.750 & 0.289 \\ 
 $\left[ 6.76 ; 21.8\right>$ Jy   &  10 & -0.790 & -0.837 & 0.335 \\ 
  \hline
 \end{tabular}  
 \end{center}
\end{table}

Sources are generally expected to follow a power law: $S(\nu)\propto\nu^{\alpha}$, where $\alpha$ is the spectral index (SI) of the source and $\nu$ the frequency. At the frequencies of interest, the average SI of sources is generally found to be around $-0.7$ to $-0.8$ \citep{intema-gmrt-bootes-I-2011, ghosh-21cm-foregrounds-2012, vanweeren-lofar-bootes-2014}. We estimate the in-band SIs of all the measured sources in our field by fitting a power law to the 40~kHz SEDs.

To assess beam-model errors, the SIs of the central 10$\degree$ of the field are plotted against the source R.A./dec. in Fig.~\ref{fig:si-vs-pos-fits}. Even in this central 10$\degree$ area, the errors in the beam model cause a significant bias of the spectral index estimates. A first order estimate of the bias is obtained by fitting the spectral index to a quadratic function over R.A. and dec., as shown by the black dashed curve in Fig.~\ref{fig:si-vs-pos-fits}. We correct the SIs for this bias, thereby keeping sources at the pointing centre constant. The applied corrections  to the SIs vary from -0.23 to 0.27. A spectral index error of 0.27 corresponds with a flux error of $\sim5$\%. This is smaller than some previous results: \citet{hurley-walker-mwacs-2014} estimate the MWA beam error to be $\sim10$\%. The improvement is mainly due to the improved MWA beam model \citep{mwa-beam-sutinjo-2015}. Nevertheless, the error is still significant, and because of this in the rest of this paper we will discard sources outside the central area of 10$\degree$ radius. This leaves 586 sources in our sample.

Beam modelling errors also cause leakage of Stokes I into the Stokes Q, U and V spectra. These errors are on the order of a few percent. We do not detect any outlying power in the Stokes Q, U and V spectra that is higher than the leakage, i.e., we do not detect any intrinsically polarized sources. We have only looked for outliers in the integrated and integrated-squared polarized flux density, no RM synthesis was performed. Besides the instrumental leakage, the ionosphere is another factor in the lack of detection of intrinsically polarized sources. Because of the long integration time, sources with linear polarization will to some extent be depolarized due to the changing total electron content (TEC) in the ionosphere, which is not taken into account in the data analysis.

\begin{figure}
\begin{center}
\includegraphics[width=8.3cm]{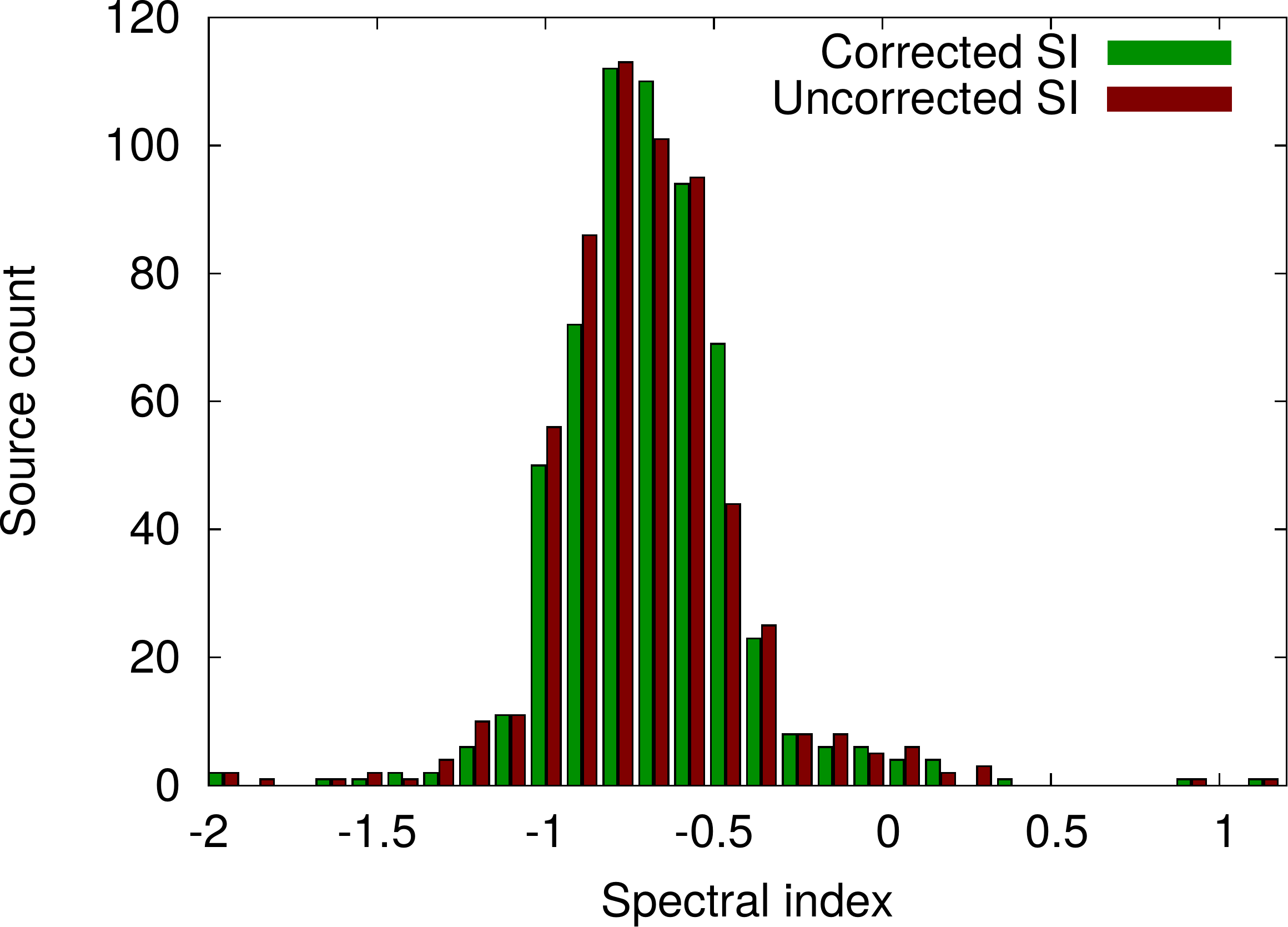}
\caption{Spectral index histogram for sources in the central $10\degree$ area, with and without correcting for beam-model errors.}
\label{fig:si-histogram}
\end{center}
\end{figure}

\begin{figure}
\begin{center}
\includegraphics[width=8.3cm]{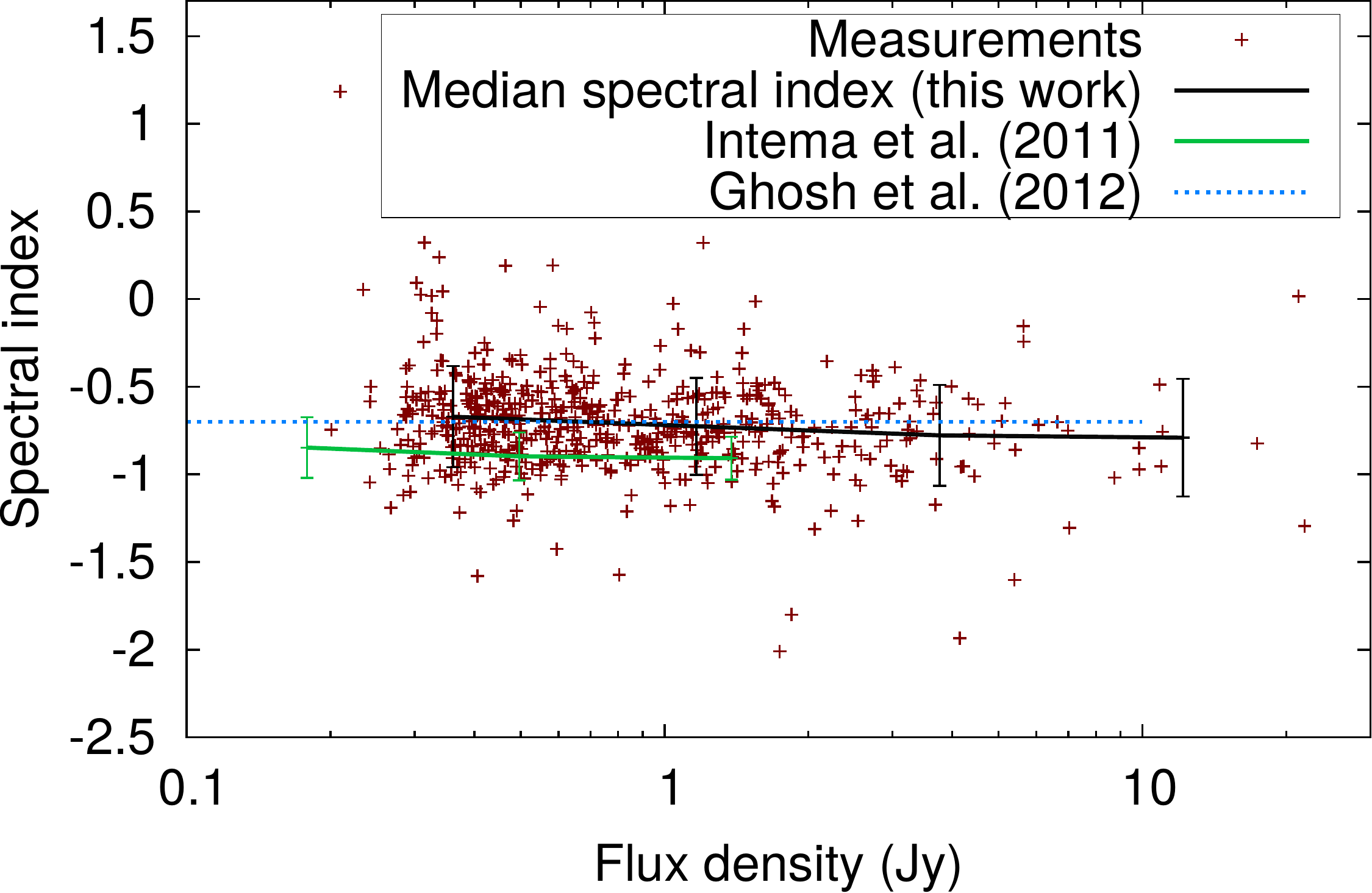}
\caption{Distribution of spectral indices over source strength. Medians were calculated by binning the sources in 4 flux density bins. The error bars indicate the standard deviation in the bin.}
\label{fig:si-trent}
\end{center}
\end{figure}

After correcting for beam errors, we find a mean $\alpha$ of -0.687, median of -0.700 and standard deviation of 0.275. A histogram is plotted in Fig.~\ref{fig:si-histogram}. We divide the observed range of flux densities in four logarithmic bins, and calculate the median spectral index for each bin. The results are shown in Table~\ref{tbl:si-bined} and plotted in Fig.~\ref{fig:si-trent}. We observe a slight but insignificant flattening towards lower flux densities. We find that the bin medians are approximately 0.15 SI units flatter compared to \citet{intema-gmrt-bootes-I-2011} at the same flux density level. Our catalogue contains $9 / 586 \approx 1.5\%$ steep spectra sources with $\alpha < -1.3$. This is a lower fraction compared to \citet{intema-gmrt-bootes-I-2011}, who find $16 / 417 \approx 3.8\%$ steep-spectrum sources.

\subsection{Spectral curvature} \label{sec:spectral-curvature}
\begin{figure}
\begin{center}
\includegraphics[width=8.3cm]{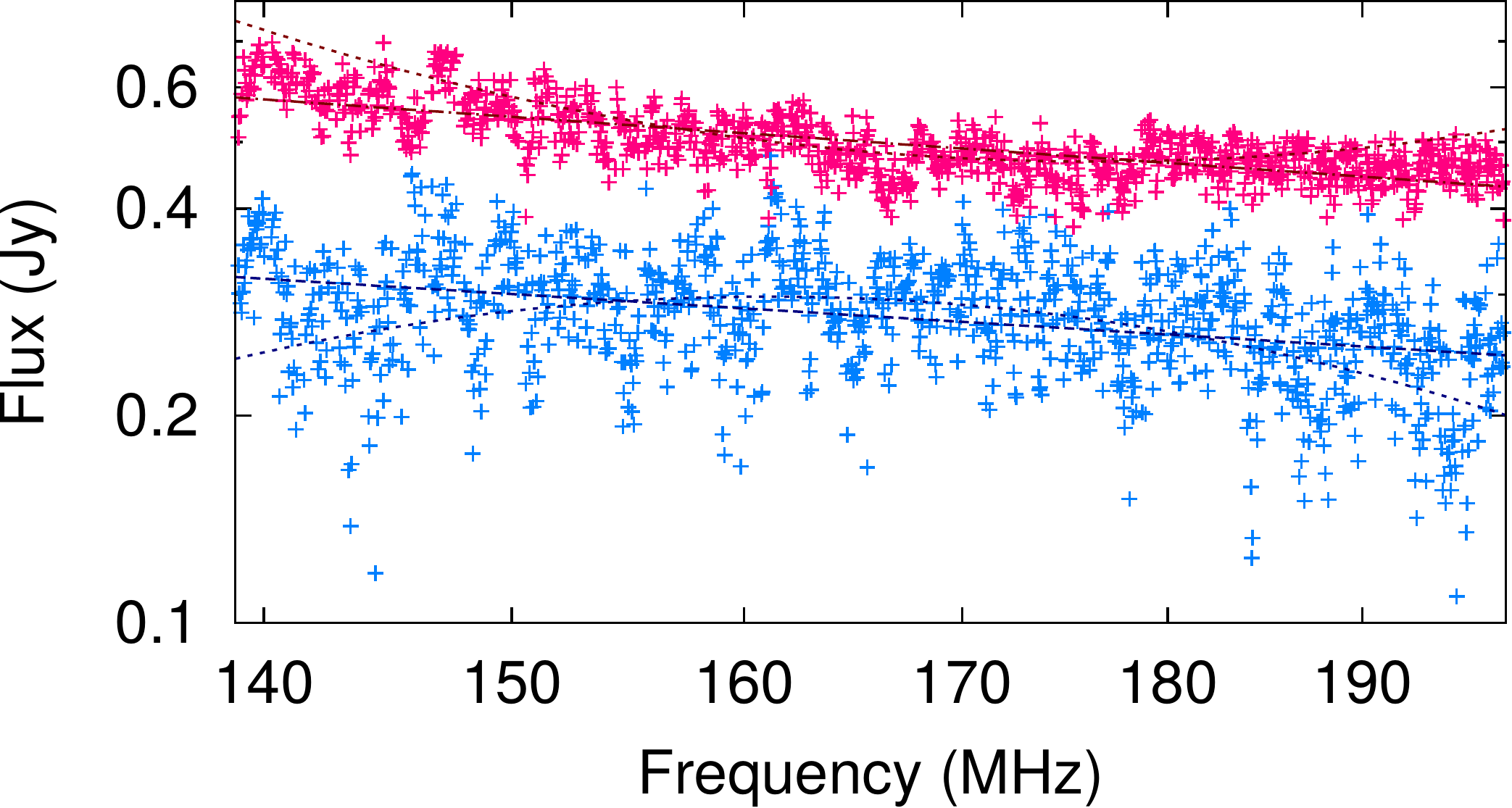}
\caption{Spectra for sources MWAEOR\,J000725-305531 (top) and MWAEOR\,J232049-254050 (bottom), which have the largest spectral curvature. The measurements are drawn together with the fitted first order (dashed lined) and second order (dotted line) fits. Clearly, the curvature resulting from fitting is not a good representation. The deviations are likely due to systematic errors and not due to instrinsic curvature of the source spectrum. These spectra have relatively large systematics when compared to other sources in the catalogue. }
\label{fig:high-curvature}
\end{center}
\end{figure}

To analyse the possible curvature of the source spectra, we fit each spectrum to a second order logarithmic polynomial:
\begin{equation} \label{eq:logpolynomial}
\log S(\nu) = \log S_0 + \alpha \log \frac{\nu}{\nu_0} + \beta \left( \log \frac{\nu}{\nu_0} \right)^2,
\end{equation}
where $S_0$ is the source flux density at the reference frequency $\nu_0=168.3$~MHz, $\alpha$ is the SI at frequency $\nu_0$ and $\beta$ is the spectral curvature. We observe that the curvature is correlated to the position of the source, similar to the spectral index. After correction for this in the same way as demonstrated in Fig.~\ref{fig:si-vs-pos-fits}, the average curvature $\hat{\mu}[\beta]=0.02$ with standard deviation $\hat{\sigma}[\beta]=2.3$. We note that the corrections are considerable: the largest absolute correction $\epsilon(\beta)$ for sources at 10$\degree$ distance is 2.1 curvature units. Although these corrections affect the mean curvature (before: 0.37, after: 0.02), its standard deviation is hardly affected (before: 2.37, after: 2.31).

When visually inspecting the spectra of the sources with largest curvature, it is apparent that most of the spectral curvature is instrumental in nature. Fig.~\ref{fig:high-curvature} shows the two spectra with the highest absolute curvature as an example. Their spectra show structure on small (few-MHz) scales, which could be caused by strong off-axis sources or the Galactic plane. Many outliers are visible as well, which could be caused by the poly-phase filter aliasing. Hence, with regards to modelling the intrinsic curvature of sources, we can only set an upper limit on its standard deviation of $\sigma[\beta] \le 2.3$. The cause of the artefacts are analysed in \S\ref{sec:cause-of-artefacts}.

\subsection{Emission / absorption line-like features} \label{sec:line-like-features}

We perform a blind search for line-like features in the spectra. The second order logarithmic polynomial from Eq.~\eqref{eq:logpolynomial} is fitted to each spectra and the maximal deviation (both positive and negative) is calculated. We calculate the significance of the deviation relative to the RMS of the difference between the model and the measured values. Initially, many deviations larger than 5$\sigma$ are found, but most of these are found to be in the first channels next to the subband edges, and caused by the poly-phase filter. After flagging a total of 6 edge channels on each side of each subband (losing 552 out of 1472 channels), 5 sources with 5$\sigma$ deviations remain. These five sources are found to have artefacts similar to Fig.~\ref{fig:high-curvature}.
With 6 sub-band edge channels removed, a maximum absolute deviation of $0.37\pm 0.22$~Jy is found in a $9.9$~Jy source, resulting in a $1.65\sigma$ deviation.
This deviation results in a $3\sigma$ upper limit of 1.03~Jy on the deviation that sources have from smooth spectra in 40~kHz channels. 

\subsection{Cause of spectral artefacts} \label{sec:cause-of-artefacts}
\begin{figure}
\begin{center}
\includegraphics[width=8.3cm]{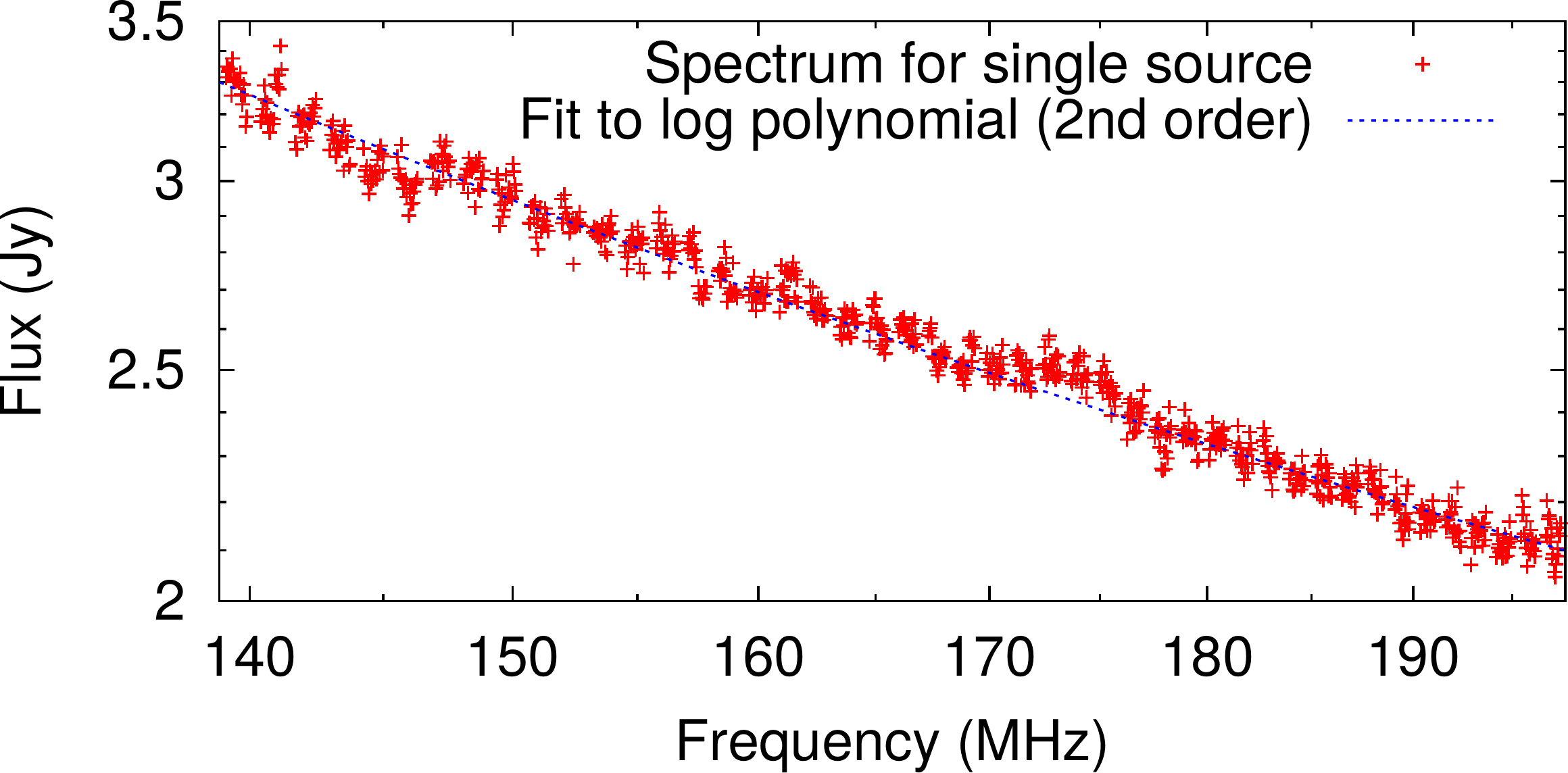}
\caption{Spectra for source MWAEOR\,J001612-312334. This source shows exceptionally large instrumental effects in its spectra. The source was found by searching for sources with large deviations from a log polynomial. Fig.~\ref{fig:cable-reflections-delays} shows the spectrum in Fourier space.}
\label{fig:cable-reflections-spectrum}
\end{center}
\end{figure}
\begin{figure}
\begin{center}
\includegraphics[width=8.5cm]{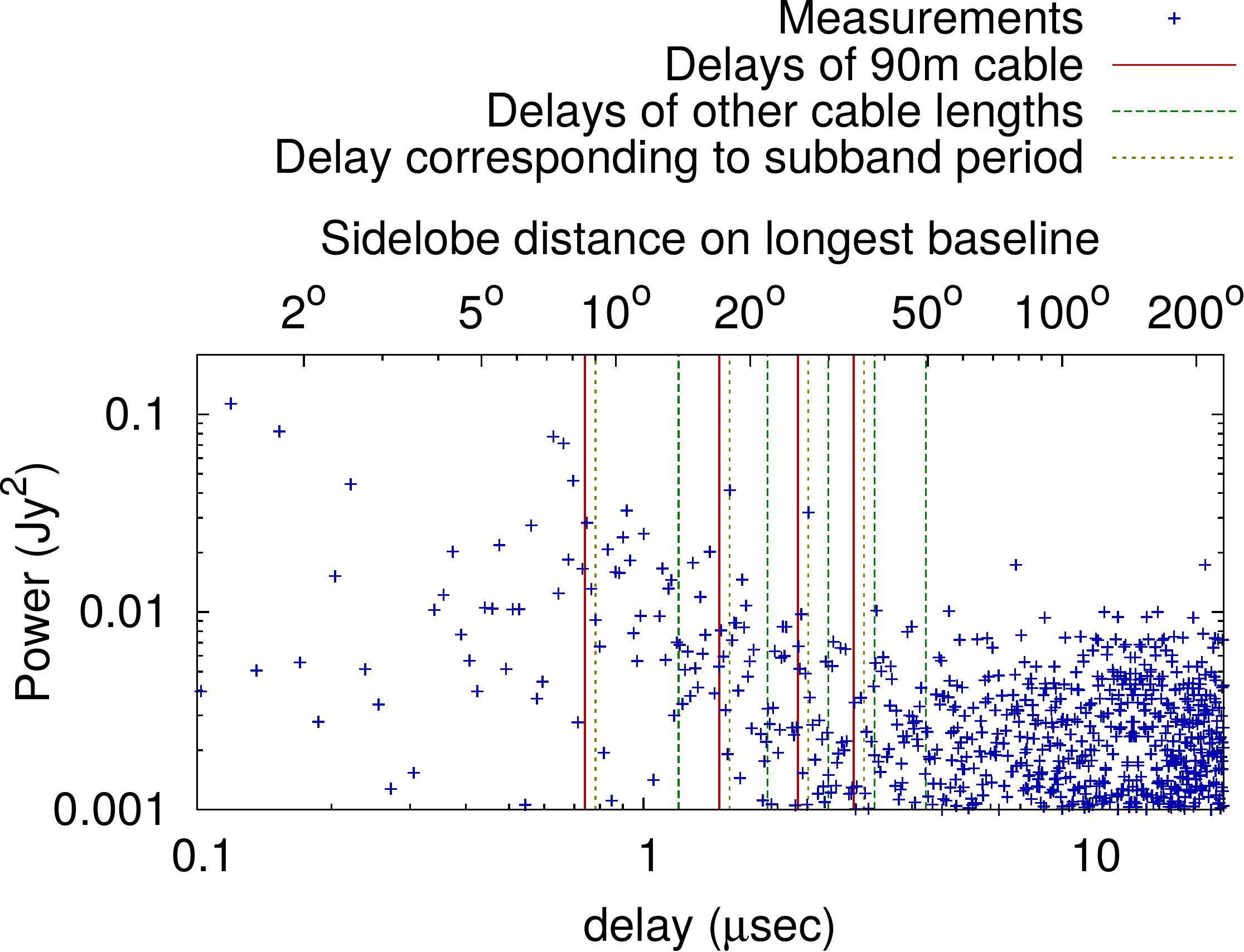}
\caption{Lomb-Scargle periodogram of the spectrum in Fig.~\ref{fig:cable-reflections-spectrum}, showing the power in delay space. Delays corresponding to cable reflections (yellow and green vertical lines) do not show increased power. The sub-band delays (red lines) do show some increased power, in particular at two and three times the corresponding delay. The sidelobe of a source can cause a fringe on the spectrum of another with a maximum fringe speed that is proportional to their distance. The top x-axis relates the delay to this distance, calculated for the longest baseline (2900 m).}
\label{fig:cable-reflections-delays}
\end{center}
\end{figure}
As described in \S\ref{sec:spectral-curvature} and \S\ref{sec:line-like-features}, the measured spectra show instrumental structures. Results of the MWA at lower frequencies have identified problems with cable reflections (Ewall-Wice et al., in prep.). Cable reflections cause a ripple over frequency, with a period that is inverse proportional to the length of the cable. To analyse whether such ripples are present in the spectra, we calculate the Lomb-Scargle periodogram for sources that show instrumental artefacts. The Lomb-Scargle periodogram is similar to the power in Fourier domain, except that missing channels from the subband edges do not cause unreal high responses at its corresponding delay \citep{scargle-1982-unevenly-spaced-periodogram}.

Fig.~\ref{fig:cable-reflections-spectrum} shows the spectrum for a source with exceptionally high artefacts, and it is one of the sources that was found to have 5$\sigma$ deviations from a smooth curve. While the spectrum shows artefacts which might appear periodic, its periodogram in Fig.~\ref{fig:cable-reflections-delays} shows that there is no excess power at delays corresponding to the cable length of 90 m or any of the other cable lengths used in the MWA. Some excess power is still seen at the second and third multiple of the subband period, which is likely due to poly-phase filter aliasing. Fig.~\ref{fig:cable-reflections-delays} is made after flagging 6 edge channels. When flagging only 2 edge channels, large power ($\sim1$ Jy$^2$) is visible at the delay corresponding to the subband period and multiples thereof. This is caused by subband aliasing.

The periodogram rules out sub-band aliasing or cable reflections as the cause of the artefacts visible in the spectrum. The self-calibration process, which finds solutions for each individual channel, has successfully removed the cable reflections, and the sub-band aliasing has been removed by extra flagging. It is therefore likely that the artefacts are caused by undeconvolved off-axis emission. This would also explain the variation of the strength of the artefacts between sources.

In a spectrum, the fringe rate of a sidelobe is linearly related to the the baseline length, as well as to the separation between the measured source and the sidelobe-inducing source. A source at 9$\degree$ distance from a measured spectrum creates at most one sidelobe per 1.28~MHz in the measured spectrum on the longest MWA baseline. Most of the excess power is at delays smaller than the delay of the sub-band bandwidth of 1.28 MHz. Some further excess is seen, but flattens at approximately a corresponding maximum distance of 20--30$\degree$. At distances larger than 30$\degree$, the power spectrum is dominated by the system noise contribution. This suggests that most of the power is coming from nearby undeconvolved sources that are within the field of view. These are therefore the sources fainter than 100~mJy that are inside the primary beam lobe, but have not been peeled. Sources outside the field of view can also add power at low delays via smaller baselines when they are in a sidelobe of the primary beam.

\subsection{Average spectrum residuals} \label{sec:average-spectrum-residuals}
\begin{figure}
\begin{center}
\includegraphics[width=8.5cm]{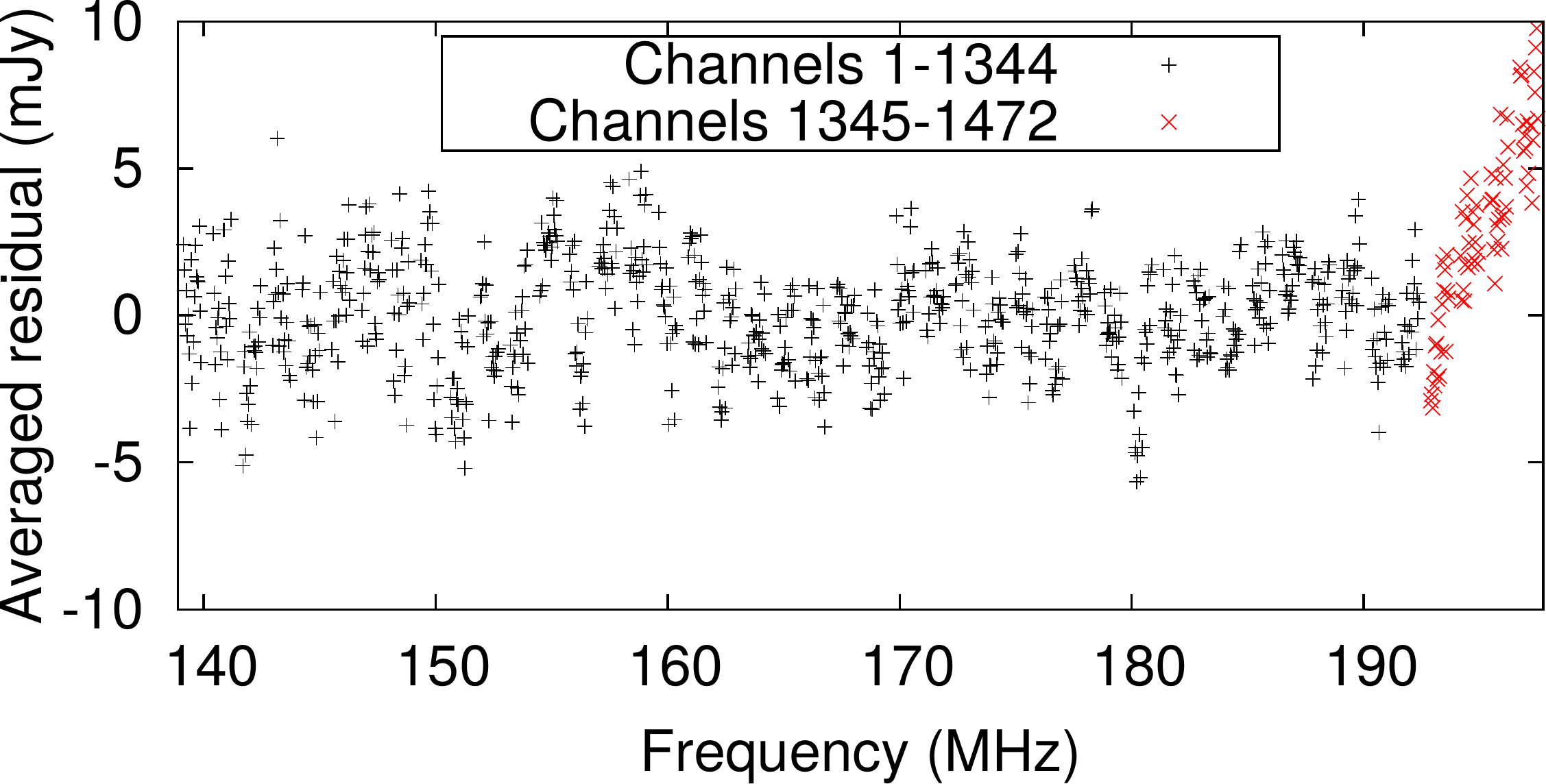}
\caption{Averaged residuals after subtracting a fitted logarithmic polynomial to each individual source spectrum. Sub-band edge channels are not plotted. Structure in the spectrum is likely due to PSF sidelobes from undeconvolved sources. The cause of the structure in channels 1345--1472 (192--198 MHz) is unknown, but most likely also caused by PSF sidelobes.}
\label{fig:average-sed-residuals}
\end{center}
\end{figure}
\begin{figure}
\begin{center}
\includegraphics[width=8.3cm]{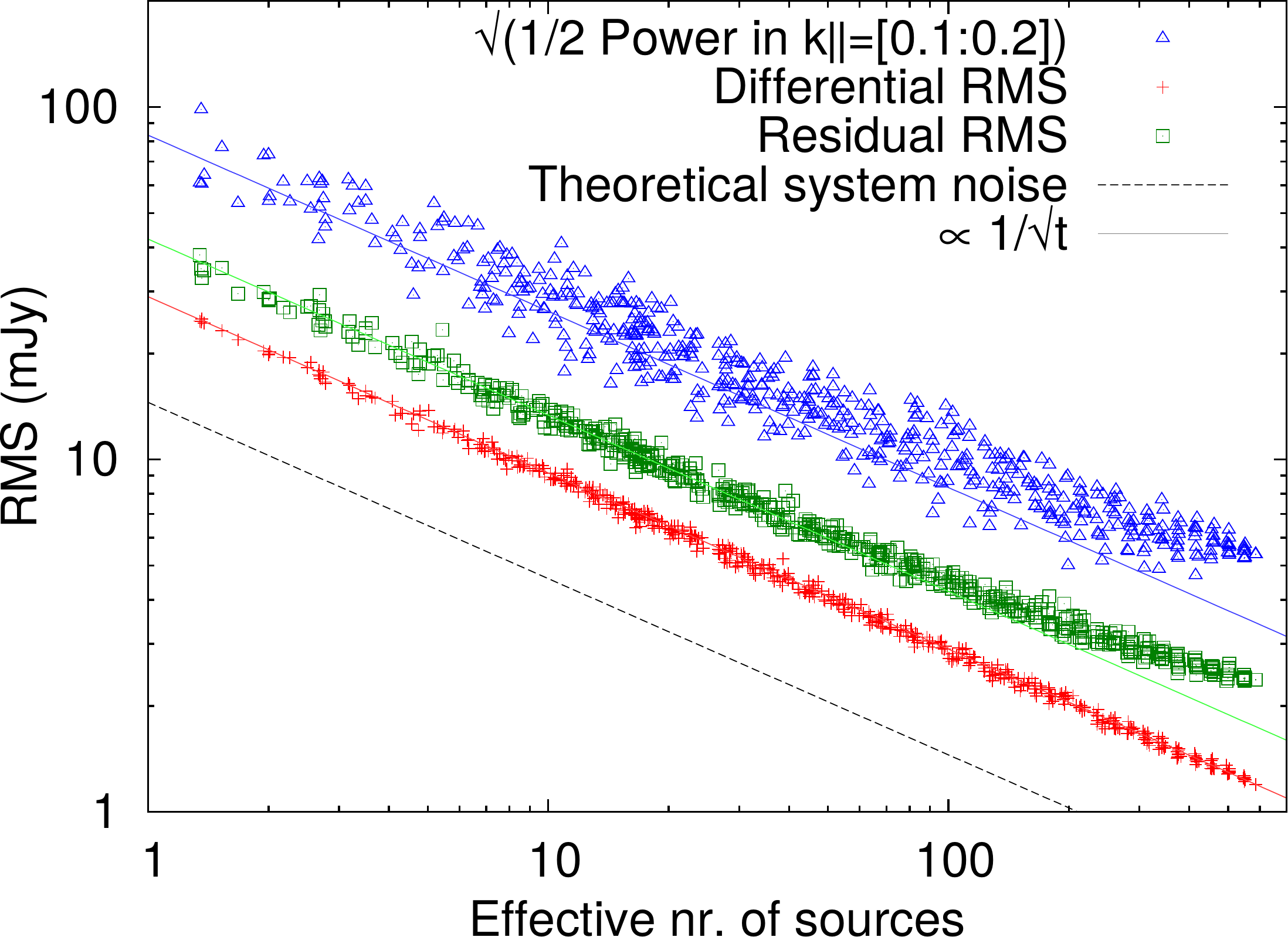}
\caption{Sensitivity of the average of spectrum residuals for different source counts, expressed with three metrics: the average power in $k_\parallel$-space range $0.1\le k_\parallel \le 0.2$ h/Mpc, the RMS of the difference between channels (differential RMS) and the normal RMS of the residuals. The theoretical system noise is corrected for uniform weighting.}
\label{fig:source-count-and-sensitivity}
\end{center}
\end{figure}
\begin{figure}
\begin{center}
\includegraphics[width=8.5cm]{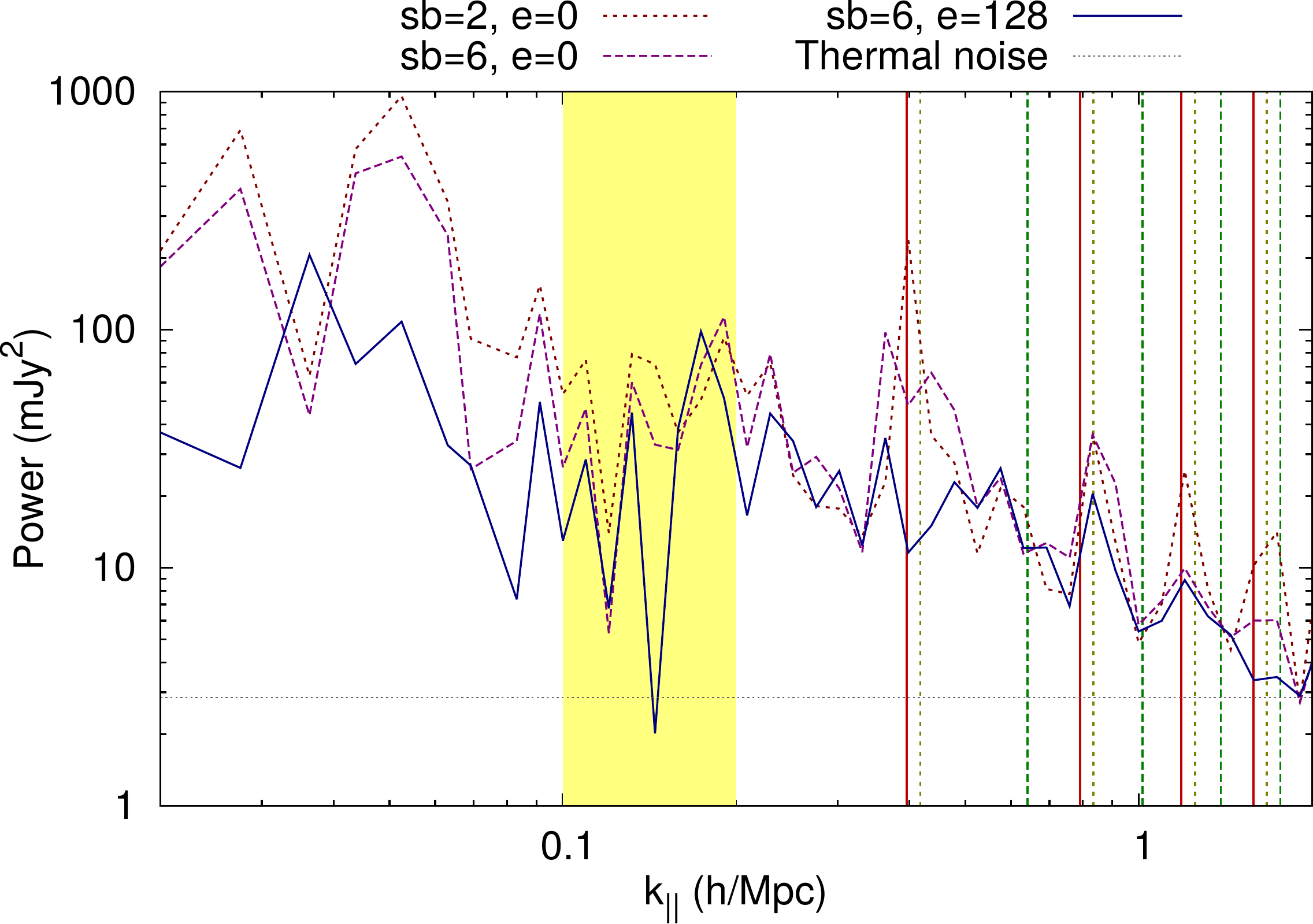}
\caption{Periodogram of the line-of-sight direction for different channel selections. The shaded area marks the range $0.1\le k_\parallel \le 0.2$ h/Mpc, which is an important part of the MWA power spectrum for the detection of EoR signals. ``sb'' and ``e'' are the number of removed sub-band edge channels and number of removed channels at the highest frequencies, respectively. The thermal noise is estimated from the differential RMS of the residuals. Vertical lines are as in Fig.~\ref{fig:cable-reflections-delays}.}
\label{fig:ps-kpar}
\end{center}
\end{figure}

So far, we have looked at the spectra of individual sources. If instrumental artefacts correlate between sources, artefacts not visible in individual source spectra might still surface after all spatial information is combined, for example by making a cylindrically-averaged or spherically-averaged power spectrum. In this section we analyse the spectral correlation between sources. We ignore the fact that sources have different positions for now, by looking at the average spectral residuals after model fitting. The presence of artefacts in averaged spectra does not strictly imply presence of artefacts in a power spectrum, but does provide an indication. In a later section we will include the spatial information by forming a circularly-averaged power spectrum.

To create an average residual spectrum, each source is individually fitted to Eq.~\ref{eq:logpolynomial} and the residuals are inverse-variance weighted before averaging. As we have already identified that sub-band edge channels are problematic, we will ignore 6 edge channels on each side of the subband. The resulting residuals are plotted in Fig.~\ref{fig:average-sed-residuals}. The structures that are visible in Figs.~\ref{fig:example-spectrum}, \ref{fig:high-curvature} and \ref{fig:cable-reflections-spectrum} are also visible in the averaged residuals, but at a smaller level. The last 128 channels ($\ge 192$ MHz) deviate, which is most likely a coincedential excess of PSF sidelobes. To quantify the artefact residuals, the RMS of the averaged residuals and the RMS of the difference between channels are calculated for different numbers of randomly selected sources.  Fig.~\ref{fig:source-count-and-sensitivity} shows the result of this. The normal RMS flattens after averaging $\sim 200$ sources. The differential RMS is not very sensitive to the structures visible in Fig.~\ref{fig:average-sed-residuals}, which is likely why the differential RMS is lower and continues to follow 1/$\sqrt{t}$ proportionally.

\subsection{$k_\parallel$ (line of sight) power spectrum}
We calculate the residual power spectrum corresponding to the residuals shown in Fig.~\ref{fig:average-sed-residuals}. This operation is equivalent to averaging lines of sight from a residual image cube, computing the power spectrum cube, and then averaging in $k_\perp$. As the average includes the wedge at scales up to the longest baselines, all modes are expected to include some measure of foreground residual. Here our intent is to highlight spectrally periodic artifacts common to all sources such as those from bandpass or reflections. As in \S\ref{sec:cause-of-artefacts}, the power   is estimated with the Lomb-Scargle periodogram. Fig.~\ref{fig:ps-kpar} shows spectra for 2 and 6 removed edge channels, and a spectrum where the deviating 6 MHz at the high-frequency end of the residuals is removed. The latter has the lowest power at almost all $k_\parallel$. When we compare this power spectrum to one calculated from a Gaussian system noise with an RMS of 1.2 mJy, which is the differential RMS of the residual spectrum, the average power in the EoR window is approximately an order of magnitude above the system noise. The power is not expected to reach the thermal noise, because the high-$k_\perp$ part of the foreground wedge is not excluded in this plot (where $k_\perp$ is the spatial direction). With only two sub-band edge channels removed, the $k_\parallel$ value that corresponds to the sub-band period shows an excess of an order of magnitude, indicating that the poly-phase filter aliasing still has a significant effect on the third sub-band edge channel. The deviating high-frequency end of the residual spectrum increases power at $k_\parallel \le 0.1$~h/Mpc in particular.

The impact of artifacts on the EoR window can be inferred by examining the power spectrum in the range of $k$-modes typically bounding the EoR window in 2D $k$-space. A comparison between the residual spectrum noise levels and the integrated power over $0.1\le k_\parallel\le 0.2$~h/Mpc for different source counts is plotted in Fig.~\ref{fig:source-count-and-sensitivity}. The power-spectrum power is scaled by plotting the square root of half the power, which implies that on average the data points would have the same positions as the RMS data points if the data are uncorrelated and Gaussian. The plot shows a larger excess for the power-spectrum power, and additionally shows that the EoR-window power flattens for high source counts similar to the RMS behaviour. PSF sidelobes from residual foregrounds are likely the cause of this. Because these statistics include high $k_\perp$-values, this is to be expected, and implies that power from the foreground wedge is contributing. When including spatial information, the power from PSF sidelobes is expected to be isolated in the wedge. If the artefacts are indeed from PSF sidelobes, we do not observe any contributions that could affect power in the EoR-window with the current sensitivity. However, some instrumental artefacts might be hard to distinguish from PSF sidelobes.

\subsection{Cylindrically-averaged two-dimensional power spectrum}
\begin{figure*}
\begin{center}
\includegraphics[height=5cm]{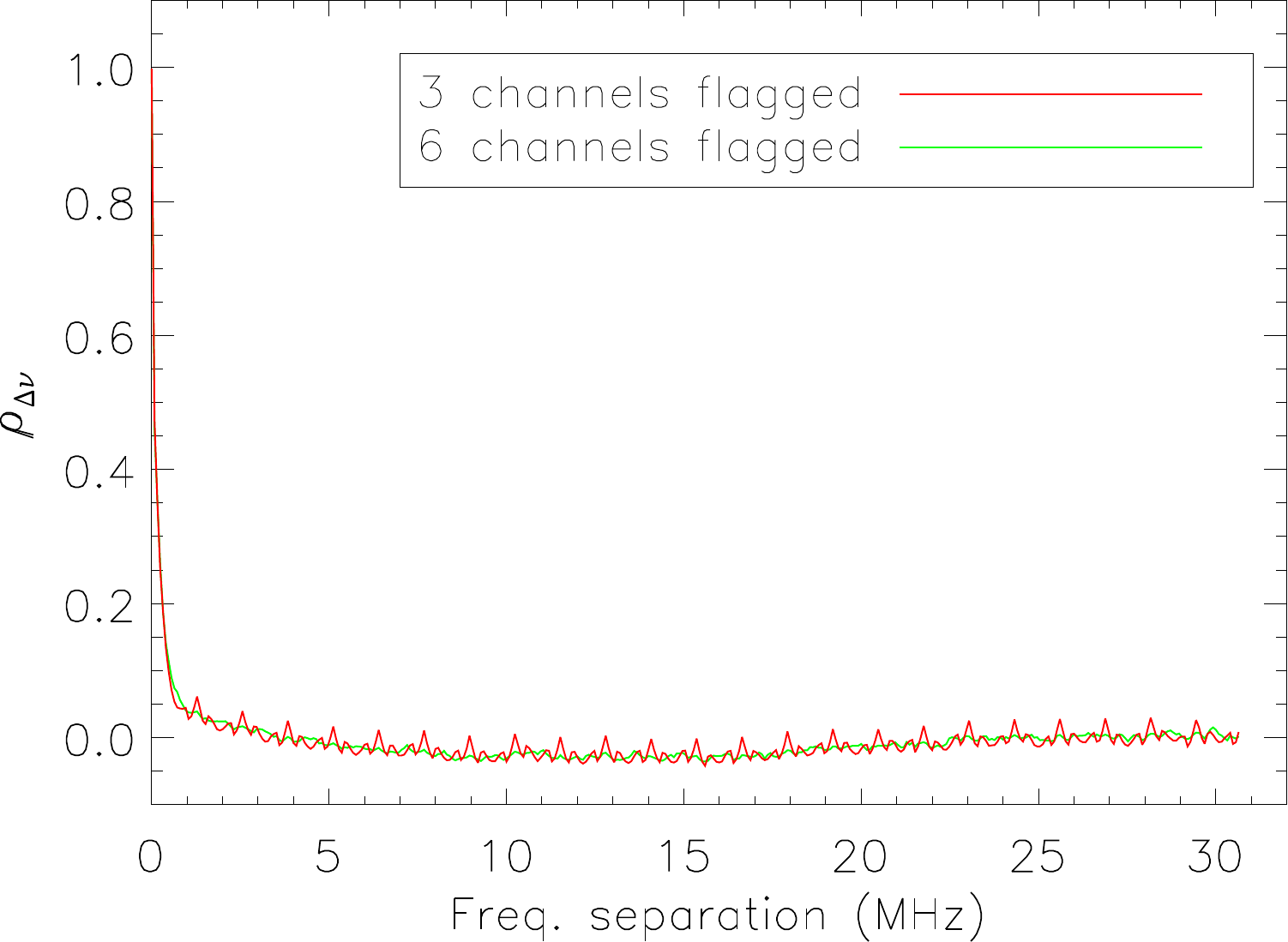}\hspace{5mm}%
\includegraphics[height=5cm]{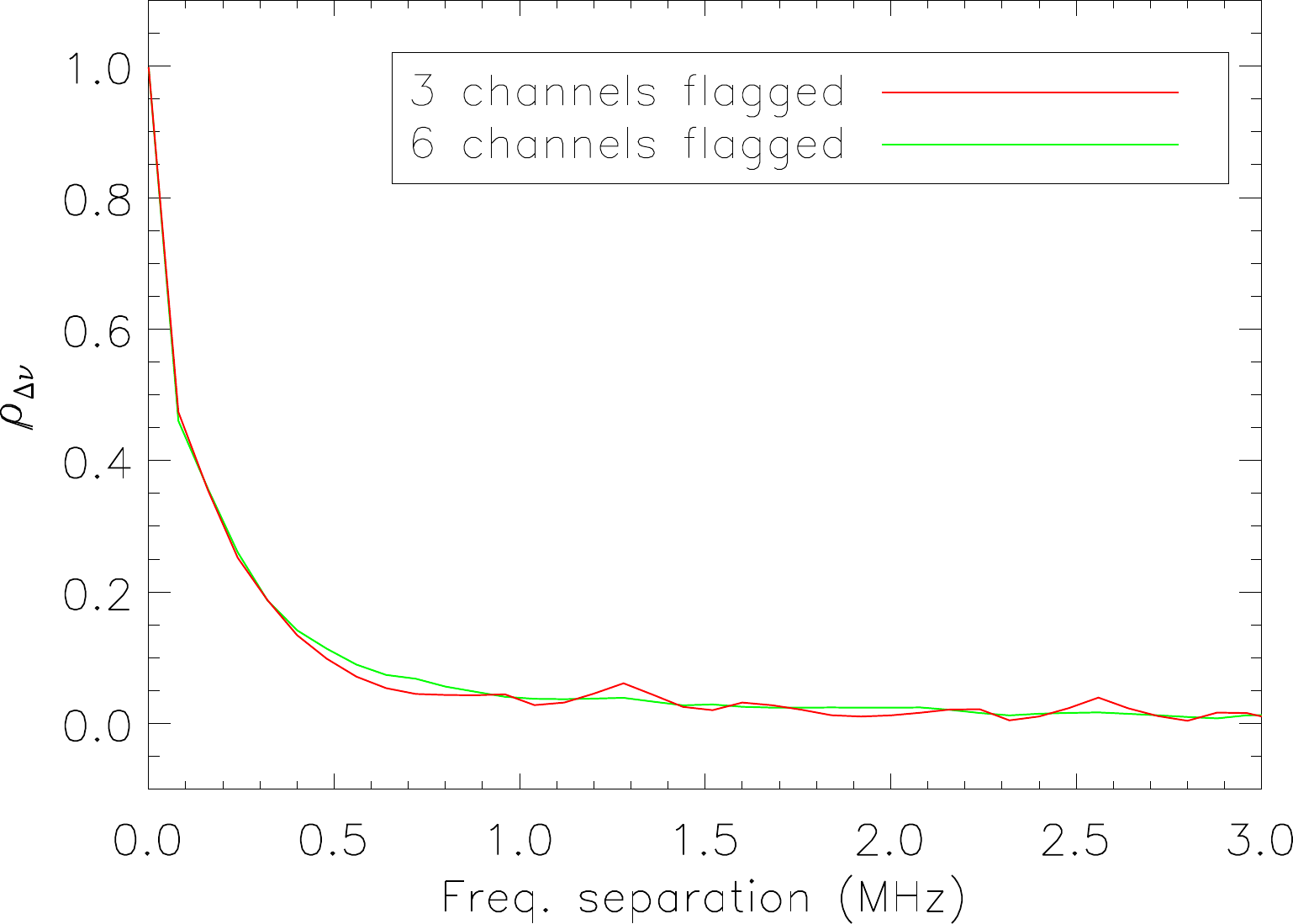}
\caption{Average two-point correlation function for all combinations of spectra, both for 3 and 6 flagged subband edge channels. The right plot is a zoom-in of the left plot.}
\label{fig:two-point-correlation}
\end{center}
\end{figure*}

\begin{figure}
\begin{center}
\includegraphics[width=8.5cm]{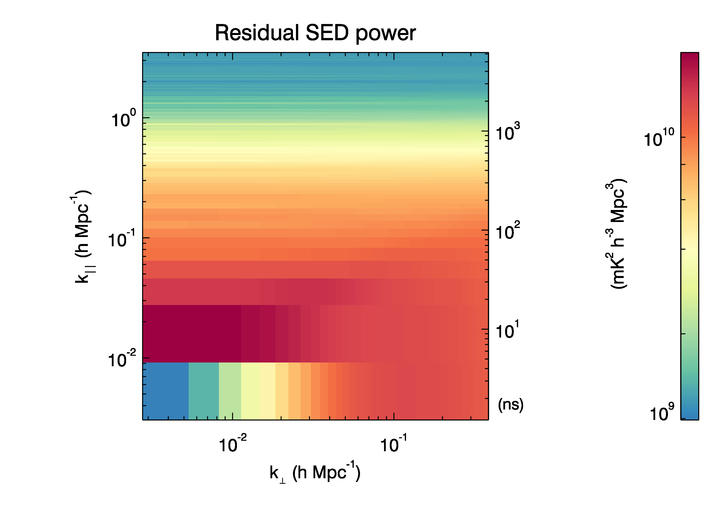}
\caption{Cylindrically-averaged power spectrum of the residual spectra, computed assuming a two-point correlation function measured from the spectral residuals, and propagated to $k_\perp$-$k_\parallel$ space, with a full instrument model.}
\label{fig:cylindrical-ps}
\end{center}
\end{figure}

\begin{figure}
\begin{center}
\includegraphics[width=8.5cm]{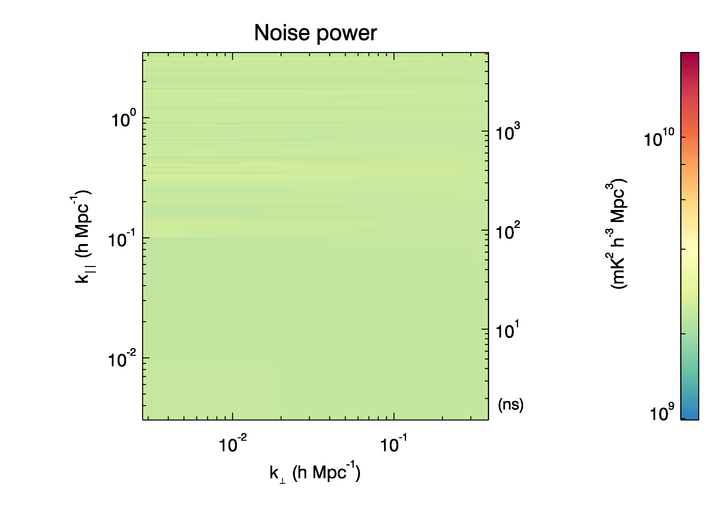}
\caption{Cylindrically-averaged power spectrum as Fig.~\ref{fig:cylindrical-ps}, but now for simulated noise.}
\label{fig:cylindrical-noise-ps}
\end{center}
\end{figure}
In addition to the simple delay-space estimate of the line-of-sight power, we compute a more sophisticated cylindrically-averaged (two-dimensional) power spectrum, including the full effects of the interferometer sampling. This analysis is based on the power spectrum estimator ``CHIPS'' developed for application to MWA EoR data, as described in \citet{trott-2015-chips}. The CHIPS estimator computes the maximum-likelihood (ML) estimate of the power. Throughout we use a $\Lambda$CDM cosmology with $H_0 = 100$h~km/s/Mpc, $\Omega_M =0.27$, $\Omega_k=0$,
$\Omega_\Lambda=0.73$ \citep{bennett-2013}.

In this analysis, the residual foregrounds are propagated into the power-spectrum parameter space, using the spectral two-point correlation function to represent the frequency-frequency covariance structure of the residual point-source spectra.

We compute the two-point correlation function,
\begin{equation}
\rho(\Delta\nu) = \frac{\left< (S(\nu_1) - \left<S\right>(\nu_1))(S(\nu_1+\Delta\nu) - \left<S\right>(\nu_1+\Delta\nu)) \right>}{\sigma(\nu_1) \sigma(\nu_1+\Delta\nu)}, 
\end{equation}
for the 586 sources using the \textit{residual} flux density (measured minus fitted) as a function of separation of spectral channels. This function is shown in Fig.~\ref{fig:two-point-correlation}. We then use the instrument chromatic sampling function, and a model for the frequency-dependent MWA primary beam, to propagate the frequency-frequency covariance of the SEDs into the power as a function of angular scale ($k_\perp$) and line-of-sight scale ($k_\parallel$). The resulting SED power spectrum is shown in Fig.~\ref{fig:cylindrical-ps}. In addition to the residual spectrum we perform the same analysis for a noise-only simulation of 586 SEDs and produce the expected noise power spectrum (Fig.~\ref{fig:cylindrical-noise-ps}). This power spectrum was made by substituting the SEDs with Gaussian simulated noise, with a standard deviation equal to the differential RMS of the SED.

The residual power spectrum here is intended to demonstrate what happens if one assumes that the spectral correlations are intrinsic to the sources. We therefore re-apply the instrument model to observe the effect. In reality, the spectral structure is due to the sidelobes from other sources, which have not been correctly accounted for due to the image-space and line-of-sight method used to compute the source SEDs.

As expected, the noise power spectrum exhibits flat power across $k_\parallel$, demonstrating the lack of frequency-frequency correlations. Conversely, the structure (correlation length) shown in the two-point correlation function is translated into a slow roll-off of power in $k_\parallel$ in the residual SED power spectrum. The other prominent features in Fig.~\ref{fig:cylindrical-ps} include (1) lower power in the DC ($k_\parallel$=0) term, attributable to the smooth power-law that has been fitted to, and subtracted from, each source spectrum; and (2) a wedge-like structure extending from low $k_\perp$--$k_\parallel$ to high $k_\perp$--$k_\parallel$, corresponding to the chromatic sampling of the interferometer (mode-mixing). This chromatic sampling would have been removed if a smooth source model was fitted over the bandpass. In that case, the sampling (which leads to the sidelobe structure in image space) would have been removed by the smoothing. However, for the case where each channel is estimated independently and the source spectrum is extracted at one position in the image, this smoothing procedure is not performed, and the resultant two-point correlation residuals retain the sidelobe structure. This is a key point for the approach and intent of this work; we are trying to extract the fine frequency information intrinsic to each source, and therefore are measuring each channel independently to probe any non-smooth spectral structure. However, in doing so, we are subject to the chromatic effects of sidelobes from nearby sources, and collect all of the undesirable instrumental effects along with any of the desired intrinsic ones. For this approach, it is difficult to disentangle these effects.

\subsection{Sensitivity analysis}
The theoretical system noise is estimated for the MWA with $t=8\times 10^4 / \left(B \sigma^2\right)$, with $\sigma$ the standard deviation in mJy/beam, $t$ the observation time in seconds and $B$ the bandwidth in MHz \citep{mwa-2013}. This formula is for natural weighting, and needs to be adapted when using uniform weighting. For the MWA, simulations of noise gridded with uniform weighting show an RMS increase of a factor of 3 with uniform weighting. The resulting estimated system noise in a single uniformly-weighted spectrum with 40~kHz resolution and 22~h integration time, is
$15 \textrm{ mJy}$. The average RMS of spectra in the inner 10$\degree$ of the primary beam is 42 mJy, and is thus a factor of 2.8 higher.
After averaging all residual spectra of the 586 sources, the RMS of the averaged spectrum is 2.4~mJy, which is a factor of 4 above the estimated system noise contribution of $600 \mu$Jy. The increase from 2.8 to 4 in the ratio between RMS and system temperature after averaging is due to artefacts that correlate between sources (\S\ref{sec:average-spectrum-residuals}). Because these artefacts are smooth, the differential RMS values are not affected. Consequently, the difference between the differential RMS and the system temperature is a factor of two both in a single residual spectrum and in the averaged residual spectrum. The factor of two difference between the system noise given by \citet{mwa-2013} and our empirical measurement of the system noise contribution in long integrations can be attributed to various practical issues, such as the loss of channels due to the passband and RFI, loss of timesteps at the beginning and end of the 2-min snapshots, bad ionospheric conditions and loss of sensitivity due to the primary beam.
 
If we assume that our differential noise levels accurately quantify the system temperature contribution, a cylindrical power spectrum made from 30~MHz of the spectra has a system noise contribution of
$2.2\times10^9$ mK$^2$h$^{-3}$Mpc$^3$, as was shown in Fig.~\ref{fig:cylindrical-noise-ps}. Of course, the power spectra derived in this work are not competitive, because they only contain information from positions on the sky at which source spectra were measured. A power spectrum that includes information from the entire field of view (within the full width at half maximum; FWHM) will be more sensitive than the power spectrum from the source spectra presented here. We can however compare the measured power spectra to the noise power spectrum with the same spatial information: Fig.~\ref{fig:cylindrical-ps} shows a 4.5 times higher power of $10^{10}$ mK$^2$ h$^{-3}$Mpc$^3$ within the EoR window $0.1 \le k_\parallel \le 0.2$ h/Mpc. Given that the only significant artefacts that we detect are PSF sidelobes, this contribution is not from intrinsic source spectra or source subtraction errors, and might be contained in the wedge when the power spectrum is directly made from a spectral cube.

We have only looked at pixels within the field of view that have the brightest sources in the MWA EoR0 field. For pixels containing bright sources, the instrumental artefacts can be higher than in quiet areas. This is the case for fitting residuals and the instrumental effects that relate to the sky brightness, such as the pass-band shape and cable reflections. After having flagged 6 channels at the edges of each subband, we found no further significant power at the delay of the passband, and cable reflections have been removed by the per-channel self-calibration process. It is therefore likely that the excess noise levels are indeed coming from PSF sidelobes of the residual foreground. Foreground sidelobes are not brighter at the positions of bright sources, hence the noise levels can be expected to hold for the entire field of view.

\section{Catalogue}
\begin{table*} \caption{The first 50 of the 586 sources in the central $10^\circ$ radius of the EoR0 field, providing a flux density measurement (in Jy), spectral index and spectral curvature for each source. A full catalogue is available on-line.} \label{tbl:catalogue-example}
\begin{tabular}{cccrrr}
\hline 
Name & RA (J2000) & Dec (J2000) & $S_\mathrm{168\,MHz}$ & $\alpha_\mathrm{168\,MHz}$ & $\beta_\mathrm{168\,MHz}$\tabularnewline
\hline 
\hline 
MWAEOR\,J000004-282420 & 00:00:04.1 & -28:24:20.2 & $0.487\pm0.028$ & $-0.67\pm0.32$ & $ 1.3\pm2.6$ \tabularnewline 
MWAEOR\,J000019-272514 & 00:00:19.4 & -27:25:14.9 & $0.360\pm0.019$ & $-0.67\pm0.29$ & $-0.6\pm2.5$ \tabularnewline 
MWAEOR\,J000027-331946 & 00:00:27.1 & -33:19:46.9 & $0.760\pm0.063$ & $-0.84\pm0.45$ & $-0.1\pm3.2$ \tabularnewline 
MWAEOR\,J000029-345223 & 00:00:29.5 & -34:52:23.9 & $1.142\pm0.103$ & $-0.75\pm0.49$ & $-1.6\pm3.3$ \tabularnewline 
MWAEOR\,J000042-342402 & 00:00:42.4 & -34:24:02.2 & $3.059\pm0.269$ & $-0.92\pm0.48$ & $-0.5\pm3.3$ \tabularnewline 
MWAEOR\,J000045-272250 & 00:00:45.6 & -27:22:50.9 & $2.089\pm0.111$ & $-0.83\pm0.29$ & $-1.1\pm2.5$ \tabularnewline 
MWAEOR\,J000046-263400 & 00:00:46.0 & -26:34:00.8 & $0.420\pm0.021$ & $-0.21\pm0.28$ & $ 0.2\pm2.5$ \tabularnewline 
MWAEOR\,J000053-355458 & 00:00:53.2 & -35:54:58.0 & $1.145\pm0.109$ & $-0.65\pm0.52$ & $-0.5\pm3.4$ \tabularnewline 
MWAEOR\,J000100-250503 & 00:01:00.0 & -25:05:03.8 & $0.942\pm0.055$ & $-0.83\pm0.32$ & $ 0.3\pm2.6$ \tabularnewline 
MWAEOR\,J000106-174126 & 00:01:06.3 & -17:41:26.9 & $1.292\pm0.123$ & $-0.36\pm0.52$ & $ 0.8\pm3.4$ \tabularnewline 
MWAEOR\,J000109-285456 & 00:01:09.2 & -28:54:56.9 & $0.444\pm0.027$ & $-0.84\pm0.33$ & $-2.6\pm2.7$ \tabularnewline 
MWAEOR\,J000117-301755 & 00:01:17.4 & -30:17:55.0 & $0.566\pm0.038$ & $-0.78\pm0.37$ & $-3.3\pm2.8$ \tabularnewline 
MWAEOR\,J000124-204005 & 00:01:24.6 & -20:40:05.9 & $0.969\pm0.078$ & $-0.64\pm0.44$ & $-0.6\pm3.1$ \tabularnewline 
MWAEOR\,J000143-305731 & 00:01:43.5 & -30:57:31.0 & $3.258\pm0.231$ & $-0.68\pm0.39$ & $-0.8\pm2.9$ \tabularnewline 
MWAEOR\,J000153-302509 & 00:01:53.4 & -30:25:09.1 & $0.653\pm0.044$ & $-0.52\pm0.37$ & $ 1.5\pm2.8$ \tabularnewline 
MWAEOR\,J000154-313936 & 00:01:54.5 & -31:39:36.0 & $0.406\pm0.030$ & $-0.62\pm0.41$ & $ 6.8\pm3.0$ \tabularnewline 
MWAEOR\,J000206-302007 & 00:02:06.5 & -30:20:07.1 & $0.727\pm0.049$ & $-0.61\pm0.37$ & $-0.2\pm2.8$ \tabularnewline 
MWAEOR\,J000211-215308 & 00:02:11.8 & -21:53:08.9 & $1.640\pm0.122$ & $-0.63\pm0.41$ & $ 0.9\pm3.0$ \tabularnewline 
MWAEOR\,J000216-282505 & 00:02:16.1 & -28:25:05.2 & $0.602\pm0.035$ & $-0.93\pm0.32$ & $ 1.3\pm2.6$ \tabularnewline 
MWAEOR\,J000218-253915 & 00:02:18.2 & -25:39:15.1 & $1.735\pm0.097$ & $-0.50\pm0.31$ & $-0.4\pm2.6$ \tabularnewline 
MWAEOR\,J000231-342613 & 00:02:31.3 & -34:26:13.9 & $0.601\pm0.053$ & $-0.79\pm0.48$ & $-3.0\pm3.3$ \tabularnewline 
MWAEOR\,J000245-302826 & 00:02:45.9 & -30:28:26.0 & $3.229\pm0.221$ & $-0.73\pm0.38$ & $-0.7\pm2.9$ \tabularnewline 
MWAEOR\,J000247-315727 & 00:02:47.2 & -31:57:27.0 & $0.289\pm0.022$ & $-0.58\pm0.42$ & $ 0.4\pm3.0$ \tabularnewline 
MWAEOR\,J000255-265451 & 00:02:55.9 & -26:54:51.1 & $0.284\pm0.015$ & $-0.67\pm0.29$ & $ 7.9\pm2.5$ \tabularnewline 
MWAEOR\,J000304-331157 & 00:03:04.2 & -33:11:57.1 & $0.505\pm0.041$ & $-0.51\pm0.45$ & $ 5.3\pm3.1$ \tabularnewline 
MWAEOR\,J000313-355634 & 00:03:13.6 & -35:56:34.1 & $5.427\pm0.520$ & $-1.47\pm0.52$ & $-1.5\pm3.4$ \tabularnewline 
MWAEOR\,J000322-172711 & 00:03:22.0 & -17:27:11.2 & $11.008\pm1.065$ & $-0.50\pm0.53$ & $-0.7\pm3.4$ \tabularnewline 
MWAEOR\,J000327-225724 & 00:03:27.5 & -22:57:24.1 & $0.828\pm0.057$ & $-0.86\pm0.38$ & $ 0.9\pm2.9$ \tabularnewline 
MWAEOR\,J000329-170631 & 00:03:29.2 & -17:06:31.0 & $0.518\pm0.051$ & $-0.85\pm0.54$ & $-0.9\pm3.5$ \tabularnewline 
MWAEOR\,J000342-213311 & 00:03:42.4 & -21:33:11.2 & $0.349\pm0.027$ & $-0.67\pm0.42$ & $-6.1\pm3.0$ \tabularnewline 
MWAEOR\,J000342-174027 & 00:03:42.5 & -17:40:27.1 & $3.682\pm0.352$ & $-0.92\pm0.52$ & $-0.5\pm3.4$ \tabularnewline 
MWAEOR\,J000348-232939 & 00:03:48.0 & -23:29:39.8 & $3.370\pm0.225$ & $-0.70\pm0.37$ & $-0.9\pm2.8$ \tabularnewline 
MWAEOR\,J000355-305953 & 00:03:55.1 & -30:59:53.2 & $5.074\pm0.361$ & $-0.67\pm0.39$ & $-0.5\pm2.9$ \tabularnewline 
MWAEOR\,J000359-270610 & 00:03:59.7 & -27:06:10.1 & $0.465\pm0.025$ & $0.20\pm0.29$ & $-1.2\pm2.5$ \tabularnewline 
MWAEOR\,J000400-263718 & 00:04:00.9 & -26:37:18.8 & $1.012\pm0.054$ & $-0.75\pm0.29$ & $ 1.5\pm2.5$ \tabularnewline 
MWAEOR\,J000402-230659 & 00:04:02.5 & -23:06:59.0 & $2.241\pm0.154$ & $-1.14\pm0.38$ & $-1.4\pm2.9$ \tabularnewline 
MWAEOR\,J000407-294010 & 00:04:07.0 & -29:40:10.9 & $0.561\pm0.036$ & $-0.74\pm0.35$ & $ 0.5\pm2.8$ \tabularnewline 
MWAEOR\,J000417-221251 & 00:04:17.1 & -22:12:51.8 & $1.651\pm0.121$ & $-0.84\pm0.40$ & $-1.8\pm3.0$ \tabularnewline 
MWAEOR\,J000421-284018 & 00:04:21.0 & -28:40:18.8 & $0.959\pm0.058$ & $-0.70\pm0.33$ & $-0.9\pm2.7$ \tabularnewline 
MWAEOR\,J000428-310753 & 00:04:28.0 & -31:07:53.0 & $0.820\pm0.059$ & $-0.39\pm0.39$ & $ 0.9\pm2.9$ \tabularnewline 
MWAEOR\,J000428-305729 & 00:04:28.3 & -30:57:29.9 & $0.543\pm0.039$ & $-0.52\pm0.39$ & $-0.1\pm2.9$ \tabularnewline 
MWAEOR\,J000453-345634 & 00:04:53.6 & -34:56:34.1 & $0.807\pm0.073$ & $-1.45\pm0.50$ & $-7.2\pm3.3$ \tabularnewline 
MWAEOR\,J000506-241313 & 00:05:06.9 & -24:13:13.1 & $0.474\pm0.030$ & $-0.55\pm0.35$ & $-1.0\pm2.8$ \tabularnewline 
MWAEOR\,J000517-183846 & 00:05:17.4 & -18:38:46.0 & $0.369\pm0.034$ & $-0.46\pm0.50$ & $-2.3\pm3.3$ \tabularnewline 
MWAEOR\,J000523-290718 & 00:05:23.5 & -29:07:18.8 & $0.357\pm0.022$ & $-0.62\pm0.34$ & $ 2.0\pm2.7$ \tabularnewline 
MWAEOR\,J000541-253853 & 00:05:41.4 & -25:38:53.9 & $0.461\pm0.027$ & $-0.51\pm0.32$ & $-3.8\pm2.6$ \tabularnewline 
MWAEOR\,J000547-193910 & 00:05:47.8 & -19:39:10.1 & $0.994\pm0.086$ & $-0.76\pm0.47$ & $ 0.9\pm3.2$ \tabularnewline 
MWAEOR\,J000553-352200 & 00:05:53.1 & -35:22:00.8 & $1.594\pm0.149$ & $-0.63\pm0.51$ & $ 0.3\pm3.4$ \tabularnewline 
MWAEOR\,J000610-343204 & 00:06:10.7 & -34:32:04.9 & $0.463\pm0.041$ & $-0.89\pm0.49$ & $ 5.1\pm3.3$ \tabularnewline 
MWAEOR\,J000636-205535 & 00:06:36.4 & -20:55:35.0 & $0.201\pm0.016$ & $-0.62\pm0.44$ & $-0.2\pm3.1$ \tabularnewline 
 &  &  &  &  & \tabularnewline
\hline 
\end{tabular}
\end{table*}
One of the results of this work is a catalogue with 586 source positions, flux density measurements, spectral indices and spectral curvature, resulting from 45 h of integration. Most of our sources are covered by existing catalogues, or will be covered by the galactic and extragalactic MWA (GLEAM) survey \citep{wayth-2015-gleam}. The source positions and flux density measurements are therefore not unique, nor do they significantly increase the accuracy of existing measurements. However, in-band spectral indices and spectral curvatures have not been available so far, and are important for EoR foreground subtraction, as well as for simulations of EoR signal extraction. The catalogue contains sources inside the central area of the field with radius 10$\degree$. The first 50 sources of the final catalogue are listed in Table~\ref{tbl:catalogue-example}. The full catalogue is available on-line.

\subsection{Error estimates}
For each source, we calculate the standard error in the flux density, spectral index and spectral curvature. We do not provide errors on the source positions, because the source positions are derived from existing surveys and not measured in this work. The contribution to the error calculations is as follows:
\begin{itemize}
 \item[-] Beam errors cause 5\% error in the flux density measurement at the edge of the catalogue area. Since the entire area is used during calibration, we add 5\% error to each source flux density measurement, and add an additional error proportional to the distance from the phase centre, adding an extra 5\% at 10$\degree$ distance.
 \item[-] We propagate the beam errors into the spectral index error, by using the fact that a 5\% error causes a spectral index error of 0.27.
 \item[-] Likewise, we propagate the beam errors into the spectral curvature by using the fact that the error is 2.1 curvature units at 10$\degree$ distance.
 \item[-] The spectral index and spectral curvature are also affected by the PSF sidelobes. Therefore, we add their measured standard deviations of 0.275 to the error in the spectral index and 2.31 to the error of each spectral curvature measurement. These values have absorbed the error contribution from the system noise, although this contribution will be small compared to instrumental effects.
\end{itemize}

Independent flux-scale errors due to errors in the catalogues used for calibration are negligible, because we have used 2500 existing sources in our calibration model.

\section{Discussion \& conclusions}
We have demonstrated several new data processing steps to extract high-resolution spectra for low-frequency radio sources. An advantage of the new approach is that it can apply ionospheric and beam-corrections at the highest time and frequency resolutions with an acceptable computational cost. Using this approach, we have reached the Stokes I confusion limit of the MWA in the integrated bandwidth, which is 3.5 mJy/beam. The differential and Stokes Q, U and V noise levels in the spectra continue to decrease with longer integration time. Our method successfully removes issues with cable reflections that have been observed at lower frequencies with the MWA.

Our measurement of the source population shows that the spectral-index distribution is similar to results with different telescopes, although our average spectral index indicates flatter spectra, with an average spectral index of $-0.69$. We also measure a larger spread compared to other studies. While the MWA might sample a slightly different source distribution because of low resolution but high sensitivity at small baselines, inaccuracies in the primary-beam model are also causing errors in the in-band spectral indices of the MWA. At 10$\degree$ from the phase centre, the spectral-index has an average error of 0.27 points caused by the primary beam model. Because of the beam errors, as well as due to the relatively high level of undeconvolved flux, we cannot measure the curvature very accurately. We have determined a spectral-curvature upper limit of $\sigma[\beta] \le 2.3$, and do not find any in-band curvature that can be confidently attributed to source-intrinsic spectral curvature. Improving the primary-beam model of the MWA is important, because it will enable more accurate in-band spectral measurements and allow using a larger area of the primary beam.

We have not found any source-intrinsic spectral lines, which rules out 40~kHz deviations $> 1.03$ Jy in our source sample. The search for these is somewhat difficult due to the poly-phase filter of the MWA as well as due to the wide field of view of the MWA. The latter requires extensive deconvolution, which is computationally expensive for a large field of view. Using the MWA to search for spectral lines in diffuse structures, such as Galactic radiation, will be more effective, because the MWA has more sensitivity and better uv-coverage at larger scales.

Due to several practical causes (loss of sub-band edges and snapshot transitions, RFI and the ionosphere) we find the effective system noise contribution to be approximately twice as high as the theoretical noise prediction that is based on the system temperature of the single elements as specified by \citet{mwa-2013}.

When flagging 3 channels on each side of each 32-channel MWA sub-band, we continue to see a large contribution in the power spectrum from the poly-phase filter. With 6 channels flagged at each side, the artefacts are mostly gone. It is possible that the poly-phase filter still contaminates the power spectrum at a fainter level, and more flagging is required for longer time integrations. A stronger poly-phase filter with a corresponding pass-band delay that does not fall in the EoR window will be advantageous for EoR experiments. This should be taken into consideration in future MWA upgrades or future telescopes such as the Square Kilometre Array (SKA).

We have looked at the power spectrum made from a limited number of source spectra. By doing so we combine the information differently compared to making power spectra from all image resolution elements (as in \citealt{mwa-eor-limit-2015}) or by making power spectra directly from the visibilities (as in \citealt{trott-2015-chips}). This different methodology has allowed us to perform extensive analysis of possible causes, but changes the contribution of certain effects somewhat.

While we observe an excess of a factor of 4.5 (in mK$^2$ h$^{-3}$Mpc$^3$) in the EoR window of the power spectrum, we conclude this is not the result of intrinsic source variation, cable reflections or pass-band ripples, but due to PSF sidelobes from unsubtracted point sources inside the primary field. It is likely that this power is mapped under the wedge when a power spectrum is made directly from a full image cube. If one would direction-dependently fit and subtract each source independently, then one would include the effect of PSF sidelobes during the subtraction, and end up with the factor 4.5 excess power. In our case, we have calibrated on clusters of sources. Such calibration strategy will decrease the effect when making a full power spectrum, but residual PSF sidelobes from unmodelled sources might still be present and affect the calibration solutions at a lower level. This shows the importance of using the best possible sky models and as little degrees of freedom as possible during calibration, because PSF sidelobes of unmodelled sources will otherwise affect the calibration solutions, and thereby propagate power to the EoR window of the power spectrum.

Making an accurate calibration model, including spectral indices and curvature, is challenging. One way to improve the results in a next iteration, is by peeling more sources and subtracting the diffuse emission from the Galaxy. Using the low and high-band images produced in this work, it is possible to construct a deeper model for this field with more accurate frequency information. Furthermore, for constraining the spectral indices and curvatures, it will be advantageous to combine data from lower and higher frequency observations. The GLEAM survey (\citealt{wayth-2015-gleam}, Hurley-Walker et al., 2016, in prep.) will provide catalogues for the MWA EoR fields. Its wider bandwidth might make it easier to construct accurate spectral indices and curvatures for the bright sources. Automated cleaning methods that incorporate spectral information, such as \textsc{casa}'s MSMFS \citep{rau-msmfs-2011} or the joined-channel cleaning methods implemented in \textsc{wsclean} and \textsc{obit} \citep{cotton-2008-obit} might be another direction worth investigating.

\section*{Acknowledgements}
AO acknowledges financial support from the European Research Council under ERC Advanced Grant LOFARCORE -- 339743. MJ-H acknowledges the support of the Marsden Fund. This research was supported under Australian Research Council's Discovery Early Career Researcher funding scheme (project number DE140100316) and the Centre for All-sky Astrophysics (an Australian Research Council Centre of Excellence funded by grant CE110001020).

This scientific work makes use of the NCI National Facility in Canberra, Australia, which is supported by the Australian Commonwealth Government.

This scientific work makes use of the Murchison Radio-astronomy Observatory, operated by CSIRO. We acknowledge the Wajarri Yamatji people as the traditional owners of the Observatory site. Support for the operation of the MWA is provided by the Australian Government Department of Industry and Science and Department of Education (National Collaborative Research Infrastructure Strategy: NCRIS), under a contract to Curtin University administered by Astronomy Australia Limited. We acknowledge the iVEC Petabyte Data Store and the Initiative in Innovative Computing and the CUDA Center for Excellence sponsored by NVIDIA at Harvard University.

\DeclareRobustCommand{\TUSSEN}[3]{#3}

\bibliographystyle{mn2e}
\bibliography{references}

\begin{thebibliography}{}

\bibitem[\protect\citeauthoryear{Ali, Bharadwaj \& Chengalur}{Ali
  et~al.}{2008}]{ali-gmrt-foregrounds-2008}
Ali S.~S.,  Bharadwaj S.,    Chengalur J.~N.,  2008, MNRAS, 385, 2166

\bibitem[\protect\citeauthoryear{Ali, Parsons, Zheng et~al.,}{Ali
  et~al.}{2015}]{ali-eor-paper-2015}
Ali Z.~S.,  Parsons A.~R.,  Zheng H.,    et~al., 2015, ApJ, 809, 61

\bibitem[\protect\citeauthoryear{Asad, Koopmans, Jeli\'c et~al.,}{Asad
  et~al.}{2015}]{asad-lofar-polarization-leakage-2015}
Asad K. M.~B.,  Koopmans L. V.~E.,  Jeli\'c V.,    et~al., 2015, MNRAS, 451,
  3709

\bibitem[\protect\citeauthoryear{Asgekar, Oonk, Yatawatta et~al.,}{Asgekar
  et~al.}{2013}]{asgekar-lofar-rrls-cas-a-2013}
Asgekar A.,  Oonk J. B.~R.,  Yatawatta S.,    et~al., 2013, A\&A, 551, L11

\bibitem[\protect\citeauthoryear{Beardsley, Hazelton, Morales
  et~al.,}{Beardsley et~al.}{2013}]{mwa-eor-sensitivity-2013}
Beardsley A.~P.,  Hazelton B.~J.,  Morales M.~F.,    et~al., 2013, MNRAS, 429,
  L5

\bibitem[\protect\citeauthoryear{Bennett, Larson, Weiland et~al.,}{Bennett
  et~al.}{2013}]{bennett-2013}
Bennett C.~L.,  Larson D.,  Weiland J.~L.,    et~al., 2013, ApJS, 208, 20

\bibitem[\protect\citeauthoryear{Bernardi, {\TUSSEN{Bruyn}{De}{de}}~Bruyn,
  Harker et~al.,}{Bernardi et~al.}{2010}]{bernardi-wsrt-foregrounds-II-2010}
Bernardi G.,  {\TUSSEN{Bruyn}{De}{de}}~Bruyn A.~G.,  Harker G.,    et~al.,
  2010, A\&A, 522, A67

\bibitem[\protect\citeauthoryear{Bernardi, McQuinn \& Greenhill}{Bernardi
  et~al.}{2015}]{bernardi-2015-leda}
Bernardi G.,  McQuinn M.,    Greenhill L.~J.,  2015, ApJ, 799, 90

\bibitem[\protect\citeauthoryear{Bowman et~al.,}{Bowman
  et~al.}{2013}]{bowman-science-with-the-mwa-2013}
Bowman J.~D.,  et~al., 2013, Pub. of the Astr. Soc. of Australia, 30, e031

\bibitem[\protect\citeauthoryear{Bowman \& Rogers}{Bowman \&
  Rogers}{2010}]{edges-2010}
Bowman J.~D.,  Rogers A. E.~E.,  2010, Nature, 468, 796

\bibitem[\protect\citeauthoryear{Burns, Lazio, Bale et~al.,}{Burns
  et~al.}{2012}]{burns-2012-dare}
Burns J.~O.,  Lazio J.,  Bale S.,    et~al., 2012, Adv. in Space Research, 49,
  433

\bibitem[\protect\citeauthoryear{{Carilli}}{{Carilli}}{1996}]{Carilli96}
{Carilli} C.~L.,  1996, A\&A, 305, 402

\bibitem[\protect\citeauthoryear{Ciardi, Labropoulos, Maselli et~al.,}{Ciardi
  et~al.}{2012}]{ciardi-2012-21cm-forest}
Ciardi B.,  Labropoulos P.,  Maselli A.,    et~al., 2012, MNRAS

\bibitem[\protect\citeauthoryear{Cotton}{Cotton}{2008}]{cotton-2008-obit}
Cotton W.~D.,  2008, PASP, 120

\bibitem[\protect\citeauthoryear{Datta, Bowman \& Carilli}{Datta
  et~al.}{2010}]{datta-2010-eor-foreground-subtraction}
Datta A.,  Bowman J.~D.,    Carilli C.~L.,  2010, ApJ, 724, 526

\bibitem[\protect\citeauthoryear{Dillon, Liu, Williams et~al.,}{Dillon
  et~al.}{2014}]{mwa32-eor-limit-2014}
Dillon J.~S.,  Liu A.,  Williams C.~L.,    et~al., 2014, Phys. Rev. D, 89,
  023002

\bibitem[\protect\citeauthoryear{Dillon, Neben, Hewitt et~al.,}{Dillon
  et~al.}{2015}]{mwa-eor-limit-2015}
Dillon J.~S.,  Neben A.~R.,  Hewitt J.~N.,    et~al., 2015, Phys. Rev. D, 91,
  123011

\bibitem[\protect\citeauthoryear{Franzen, Jackson, Offringa et~al.,}{Franzen
  et~al.}{Submitted}]{franzen-2015-mwa-source-population}
Franzen T.,  Jackson C.~A.,  Offringa A.~R.,    et~al., Submitted, MNRAS

\bibitem[\protect\citeauthoryear{Garn, Green, Hales, Riley \& Alexander}{Garn
  et~al.}{2007}]{garn-2007-gmrt-eg-survey}
Garn T.,  Green D.~A.,  Hales S. E.~G.,  Riley J.~M.,    Alexander P.,  2007,
  MNRAS, 376, 1251

\bibitem[\protect\citeauthoryear{Geil, Gaensler \& Wyithe}{Geil
  et~al.}{2011}]{geil-2011-pol-foregrounds}
Geil P.~M.,  Gaensler B.~M.,    Wyithe J. S.~B.,  2011, MNRAS, 418, 516

\bibitem[\protect\citeauthoryear{Ghosh, Prasad, Bharadwaj, Ali \&
  Chengalur}{Ghosh et~al.}{2012}]{ghosh-21cm-foregrounds-2012}
Ghosh A.,  Prasad J.,  Bharadwaj S.,  Ali S.~S.,    Chengalur J.~N.,  2012,
  MNRAS, 426, 3295

\bibitem[\protect\citeauthoryear{Hancock, Murphy, Gaensler, Hopkins \&
  Curran}{Hancock et~al.}{2012}]{aegean-hancock-2012}
Hancock P.~J.,  Murphy T.,  Gaensler B.~M.,  Hopkins A.,    Curran J.~R.,
  2012, MNRAS, 422, 1812

\bibitem[\protect\citeauthoryear{Heald, Pizzo, Orr\`u et~al.,}{Heald
  et~al.}{2015}]{heald-2015-msss}
Heald G.~H.,  Pizzo R.~F.,  Orr\`u E.,    et~al., 2015, A\&A, 582, A123

\bibitem[\protect\citeauthoryear{Heesen, Beck, Krause \& Dettmar}{Heesen
  et~al.}{2011}]{Heesen11}
Heesen V.,  Beck R.,  Krause M.,    Dettmar R.-J.,  2011, A\&A, 535, A79

\bibitem[\protect\citeauthoryear{Heesen, Krause, Beck \& Dettmar}{Heesen
  et~al.}{2005}]{Heesen05}
Heesen V.,  Krause M.,  Beck R.,    Dettmar R.,  2005, in {Chyzy} K.~T.,
  {Otmianowska-Mazur} K.,  {Soida} M.,   {Dettmar} R.-J.,  eds, The Magnetized
  Plasma in Galaxy Evolution Vol.~783, {The Radio Halo of the Starburst Galaxy
  NGC 253}.
Bad Honnef (Germany), pp 336--342

\bibitem[\protect\citeauthoryear{Hurley-Walker, Morgan, Wayth
  et~al.,}{Hurley-Walker et~al.}{2014}]{hurley-walker-mwacs-2014}
Hurley-Walker N.,  Morgan J.,  Wayth R.~B.,    et~al., 2014, PASA, 31

\bibitem[\protect\citeauthoryear{Intema, van Weeren, R\"ottgering \&
  Lal}{Intema et~al.}{2011}]{intema-gmrt-bootes-I-2011}
Intema H.~T.,  van Weeren R.~J.,  R\"ottgering H. J.~A.,    Lal D.~V.,  2011,
  A\&A, 535, A38

\bibitem[\protect\citeauthoryear{Jacobs, Aguirre, Parsons, Pober, Bradley,
  Carilli, Gugliucci, Manley, van~der Merwe, Moore \& Parashare}{Jacobs
  et~al.}{2011}]{jacobs-2011-paper-survey}
Jacobs D.~C.,  Aguirre J.~E.,  Parsons A.~R.,  Pober J.~C.,  Bradley R.~F.,
  Carilli C.,  Gugliucci N.~E.,  Manley J.~R.,  van~der Merwe C.,  Moore D.~F.,
     Parashare C.,  2011, ApJ, 734, L34

\bibitem[\protect\citeauthoryear{Jeli\'c, {\TUSSEN{Bruyn}{De}{de}}~Bruyn,
  Mevius et~al.,}{Jeli\'c et~al.}{2014}]{jelic-lofar-foregrounds-II-2014}
Jeli\'c V.,  {\TUSSEN{Bruyn}{De}{de}}~Bruyn A.~G.,  Mevius M.,    et~al., 2014,
  A\&A, 568, A101

\bibitem[\protect\citeauthoryear{Jeli\'c, Zaroubi, Labropoulos, Thomas,
  Bernardi, Brentjens, De~Bruyn, Ciardi, Harker, Koopmans, Pandey, Schaye \&
  Yatawatta}{Jeli\'c et~al.}{2008}]{jelic-lofar-foregrounds-2008}
Jeli\'c V.,  Zaroubi S.,  Labropoulos P.,  Thomas R.~M.,  Bernardi G.,
  Brentjens M.~A.,  De~Bruyn A.~G.,  Ciardi B.,  Harker G.,  Koopmans L. V.~E.,
   Pandey V.~N.,  Schaye J.,    Yatawatta S.,  2008, MNRAS, 389, 1319

\bibitem[\protect\citeauthoryear{{Karachentsev}}{{Karachentsev}}{2005}]{Karachentsev05}
{Karachentsev} I.~D.,  2005, AJ, 129, 178

\bibitem[\protect\citeauthoryear{Kazemi, Yatawatta \& Zaroubi}{Kazemi
  et~al.}{2013}]{kazemi-clustered-calibration-2013}
Kazemi S.,  Yatawatta S.,    Zaroubi S.,  2013, MNRAS, 430, 1457

\bibitem[\protect\citeauthoryear{Large, Mills, Little, Crawford \&
  Sutton}{Large et~al.}{1981}]{mrc-1981}
Large M.,  Mills B.,  Little A.,  Crawford D.,    Sutton J.,  1981, MNRAS, 194,
  693

\bibitem[\protect\citeauthoryear{{Lenc} \& {Tingay}}{{Lenc} \&
  {Tingay}}{2006}]{Lenc06}
{Lenc} E.,  {Tingay} S.~J.,  2006, AJ, 132, 1333

\bibitem[\protect\citeauthoryear{Liu, Tegmark \& Zaldarriaga}{Liu
  et~al.}{2009}]{liu-2009}
Liu A.,  Tegmark M.,    Zaldarriaga M.,  2009, MNRAS, 394, 1575

\bibitem[\protect\citeauthoryear{McQuinn, Zahn, Zaldarriaga, Hernquist \&
  Furlanetto}{McQuinn et~al.}{2006}]{mcquinn-2006}
McQuinn M.,  Zahn O.,  Zaldarriaga M.,  Hernquist L.,    Furlanetto S.~R.,
  2006, The Astrophysical Journal, 653, 815

\bibitem[\protect\citeauthoryear{Mauch, Murphy, Buttery et~al.,}{Mauch
  et~al.}{2003}]{summs-II-2003}
Mauch T.,  Murphy T.,  Buttery H.~J.,    et~al., 2003, MNRAS, 342, 1117

\bibitem[\protect\citeauthoryear{Mitchell, Greenhill, Wayth, Sault, Lonsdale,
  Cappallo, Morales \& Ord}{Mitchell et~al.}{2008}]{rts-mwa-2008}
Mitchell D.,  Greenhill L.,  Wayth R.,  Sault R.,  Lonsdale C.,  Cappallo R.,
  Morales M.,    Ord S.,  2008, Selected Topics in Sig. Proc., IEEE Journal of,
  2, 707

\bibitem[\protect\citeauthoryear{Morabito, Oonk, Salgado et~al.,}{Morabito
  et~al.}{2014}]{morabito-2014}
Morabito L.~K.,  Oonk J. B.~R.,  Salgado F.,    et~al., 2014, ApJ, 795, L33

\bibitem[\protect\citeauthoryear{Morales, Hazelton, Sullivan \&
  Beardsley}{Morales et~al.}{2012}]{morales-2012-eorwindow}
Morales M.~F.,  Hazelton B.,  Sullivan I.,    Beardsley A.,  2012, ApJ, 752,
  137

\bibitem[\protect\citeauthoryear{Motch, Pakull, Soria, Gris{\'e} \&
  Pietrzy{\'n}ski}{Motch et~al.}{2014}]{Motch14}
Motch C.,  Pakull M.~W.,  Soria R.,  Gris{\'e} F.,    Pietrzy{\'n}ski G.,
  2014, Nature, 514, 198

\bibitem[\protect\citeauthoryear{Offringa, {\TUSSEN{Bruyn}{De}{de}}~Bruyn,
  Biehl, Zaroubi, Bernardi \& Pandey}{Offringa
  et~al.}{2010}]{offringa-2010-post-correlation-rfi-classification}
Offringa A.~R.,  {\TUSSEN{Bruyn}{De}{de}}~Bruyn A.~G.,  Biehl M.,  Zaroubi S.,
  Bernardi G.,    Pandey V.~N.,  2010, MNRAS, 405, 155

\bibitem[\protect\citeauthoryear{Offringa, {\TUSSEN{Gronde}{Van}{van}}~de
  Gronde \& Roerdink}{Offringa
  et~al.}{2012}]{offringa-2012-scale-invariant-rank-operator}
Offringa A.~R.,  {\TUSSEN{Gronde}{Van}{van}}~de Gronde J.~J.,    Roerdink J. B.
  T.~M.,  2012, A\&A, 539

\bibitem[\protect\citeauthoryear{Offringa, McKinley, Hurley-Walker
  et~al.,}{Offringa et~al.}{2014}]{offringa-wsclean-2014}
Offringa A.~R.,  McKinley B.,  Hurley-Walker N.,    et~al., 2014, MNRAS, 444,
  606

\bibitem[\protect\citeauthoryear{Offringa, Wayth, Hurley-Walker
  et~al.,}{Offringa et~al.}{2015}]{offringa-2015-mwa-radio-environment}
Offringa A.~R.,  Wayth R.~B.,  Hurley-Walker N.,    et~al., 2015, PASA, 32

\bibitem[\protect\citeauthoryear{Paciga, Albert, Bandura et~al.,}{Paciga
  et~al.}{2013}]{paciga-2013-GMRT-EoR}
Paciga G.,  Albert J.~G.,  Bandura K.,    et~al., 2013, MNRAS, 433, 639

\bibitem[\protect\citeauthoryear{{Pannuti}, {Duric}, {Lacey}, {Ferguson},
  {Magnor} \& {Mendelowitz}}{{Pannuti} et~al.}{2002}]{Pannuti02}
{Pannuti} T.~G.,  {Duric} N.,  {Lacey} C.~K.,  {Ferguson} A.~M.~N.,  {Magnor}
  M.~A.,    {Mendelowitz} C.,  2002, ApJ, 565, 966

\bibitem[\protect\citeauthoryear{Parsons, Pober, Aguirre, Carilli, Jacobs \&
  Moore}{Parsons et~al.}{2012}]{parsons-2012-delay-spectrum}
Parsons A.~R.,  Pober J.~C.,  Aguirre J.~E.,  Carilli C.~L.,  Jacobs D.~C.,
  Moore D.~F.,  2012, ApJ, 756, 165

\bibitem[\protect\citeauthoryear{Rampadarath, Morgan, Lenc \&
  Tingay}{Rampadarath et~al.}{2014}]{rampadarath-2014-ngc253}
Rampadarath H.,  Morgan J.~S.,  Lenc E.,    Tingay S.~J.,  2014, AJ, 147, 5

\bibitem[\protect\citeauthoryear{Rau \& Cornwell}{Rau \&
  Cornwell}{2011}]{rau-msmfs-2011}
Rau U.,  Cornwell T.~J.,  2011, A\&A, 532, A71

\bibitem[\protect\citeauthoryear{Salvini \& Wijnholds}{Salvini \&
  Wijnholds}{2014}]{stefcal-2014}
Salvini S.,  Wijnholds S.,  2014, in General Assembly and Scientific Symposium
  (URSI GASS), 2014 XXXIth URSI Stefcal --- an alternating direction implicit
  method for fast full polarization array calibration.
pp~1--4

\bibitem[\protect\citeauthoryear{Scargle}{Scargle}{1982}]{scargle-1982-unevenly-spaced-periodogram}
Scargle J.~D.,  1982, ApJ, 263, 835

\bibitem[\protect\citeauthoryear{Slee}{Slee}{1995}]{calgoora-1995}
Slee O.~B.,  1995, Aust. J. Phys., 48, 143

\bibitem[\protect\citeauthoryear{Sokolowski, Tremblay, Wayth
  et~al.,}{Sokolowski et~al.}{2015}]{bighorns-2015}
Sokolowski M.,  Tremblay S.~E.,  Wayth R.~B.,    et~al., 2015, PASA, 32

\bibitem[\protect\citeauthoryear{Sutinjo, O'Sullivan, Lenc, Wayth, Padhi, Hall
  \& Tingay}{Sutinjo et~al.}{2015}]{mwa-beam-sutinjo-2015}
Sutinjo A.,  O'Sullivan J.,  Lenc E.,  Wayth R.~B.,  Padhi S.,  Hall P.,
  Tingay S.~J.,  2015, RS, 50, 52

\bibitem[\protect\citeauthoryear{Thyagarajan, Jacobs, Bowman
  et~al.,}{Thyagarajan et~al.}{2015}]{thyagarajan-2015-widefield-effect}
Thyagarajan N.,  Jacobs D.~C.,  Bowman J.~D.,    et~al., 2015, ApJ, 807, L28

\bibitem[\protect\citeauthoryear{Thyagarajan, Shankar, Subrahmanyan
  et~al.,}{Thyagarajan et~al.}{2013}]{thyagarajan-2013-mwa-eor-foregrounds}
Thyagarajan N.,  Shankar N.~U.,  Subrahmanyan R.,    et~al., 2013, ApJ, 776, 6

\bibitem[\protect\citeauthoryear{{Tingay}}{{Tingay}}{2004}]{Tingay04}
{Tingay} S.~J.,  2004, AJ, 127, 10

\bibitem[\protect\citeauthoryear{Tingay et~al.,}{Tingay
  et~al.}{2013}]{mwa-2013}
Tingay S.~J.,  et~al., 2013, PASA, 30, e007

\bibitem[\protect\citeauthoryear{Tremblay, Walsh, Hurley-Walker, Wayth \&
  Hancock}{Tremblay et~al.}{submitted}]{tremblay-2015-mwa-molecular-lines}
Tremblay C.~D.,  Walsh A.~J.,  Hurley-Walker N.,  Wayth R.,    Hancock P.~J.,
  submitted, MNRAS

\bibitem[\protect\citeauthoryear{Trott et~al.,}{Trott
  et~al.}{submitted}]{trott-2015-chips}
Trott C.~M.,  et~al., submitted, ApJ

\bibitem[\protect\citeauthoryear{Trott, Wayth \& Tingay}{Trott
  et~al.}{2012}]{trott-2012-eor-point-sources}
Trott C.~M.,  Wayth R.~B.,    Tingay S.~J.,  2012, ApJ, 757, 101

\bibitem[\protect\citeauthoryear{{Turner} \& {Ho}}{{Turner} \&
  {Ho}}{1985}]{Turner85}
{Turner} J.~L.,  {Ho} P.~T.~P.,  1985, ApJ, 299, L77

\bibitem[\protect\citeauthoryear{{Ulvestad} \& {Antonucci}}{{Ulvestad} \&
  {Antonucci}}{1997}]{Ulvestad97}
{Ulvestad} J.~S.,  {Antonucci} R.~R.~J.,  1997, ApJ, 488, 621

\bibitem[\protect\citeauthoryear{van Weeren, Williams, Tasse et~al.,}{van
  Weeren et~al.}{2014}]{vanweeren-lofar-bootes-2014}
van Weeren R.~J.,  Williams W.~L.,  Tasse C.,    et~al., 2014, ApJ, 793, 82

\bibitem[\protect\citeauthoryear{Vedantham, Shankar \& Subrahmanyan}{Vedantham
  et~al.}{2012}]{vedantham-eor-foregrounds-2012}
Vedantham H.,  Shankar N.~U.,    Subrahmanyan R.,  2012, ApJ, 745, 176

\bibitem[\protect\citeauthoryear{Voytek, Natarajan, J\'auregui~Garc\'ia,
  Peterson \& L\'opez-Cruz}{Voytek et~al.}{2014}]{voytek-2014-scihi}
Voytek T.~C.,  Natarajan A.,  J\'auregui~Garc\'ia J.~M.,  Peterson J.~B.,
  L\'opez-Cruz O.,  2014, ApJ, 782, L9

\bibitem[\protect\citeauthoryear{Wang, Tegmark, Santos \& Knox}{Wang
  et~al.}{2006}]{wang-2006}
Wang X.,  Tegmark M.,  Santos M.~G.,    Knox L.,  2006, ApJ, 650, 529

\bibitem[\protect\citeauthoryear{Wayth, Lenc, Bell et~al.,}{Wayth
  et~al.}{2015}]{wayth-2015-gleam}
Wayth R.~B.,  Lenc E.,  Bell M.~E.,    et~al., 2015, PASA, 32

\bibitem[\protect\citeauthoryear{Williams, Intema \& Röttgering}{Williams
  et~al.}{2013}]{williams-gmrt-mini-survey-2013}
Williams W.~L.,  Intema H.~T.,    Röttgering H. J.~A.,  2013, A\&A, 549, A55

\bibitem[\protect\citeauthoryear{Yatawatta, de Bruyn, Brentjens
  et~al.,}{Yatawatta et~al.}{2013}]{ncp-eor-yatawatta}
Yatawatta S.,  de Bruyn A.~G.,  Brentjens M.~A.,    et~al., 2013, A\&A, 550

\end{thebibliography}

\label{lastpage}

\end{document}